\documentclass[letterpaper, 10 pt, journal, twoside]{IEEEtran}

\IEEEoverridecommandlockouts

{}
{}
{}

\usepackage{bm}

\usepackage{algorithmic}

\usepackage{amsmath,amssymb,amsfonts}
\usepackage{graphicx}
\usepackage{textcomp}
\usepackage{xcolor}
\usepackage{tabularray}
\usepackage{multirow}
\usepackage{caption}
\usepackage{subcaption}

\usepackage{capt-of}
\usepackage{cuted}
\usepackage{lipsum}

\usepackage[linesnumbered]{algorithm2e}

\def\BibTeX{{\rm B\kern-.05em{\sc i\kern-.025em b}\kern-.08em
    T\kern-.1667em\lower.7ex\hbox{E}\kern-.125emX}}

\title{A Reinforcement Learning Engine with Reduced Action and State Space for Scalable Cyber-Physical Optimal 
Response
\\ 
}

\author{Shining~Sun$^{1}$, Khandaker Akramul~Haque$^{1}$, Xiang Huo$^{1}$, Leen~Al Homoud$^{1}$, Shamina~Hossain-McKenzie$^{2}$, Ana~Goulart$^{1}$, Katherine~Davis$^{1}$
\thanks{Shining~Sun$^{1}$, Khandaker Akramul~Haque$^{1}$, Xiang Huo$^{1}$, Leen~Al Homoud$^{1}$, Ana~Goulart$^{1}$, and  Katherine~Davis$^{1}$ are with the Department of Electrical and Computer Engineering, Texas A\&M University, College Station, TX, USA; email: {xiang.huo@tamu.edu}. Shamina~Hossain-McKenzie$^{2}$ is with the Sandia National Laboratories, Albuquerque, NM, USA.}
}

\begin{document}

\maketitle

\begin{abstract}

Numerous research studies have been conducted to enhance the resilience of cyber-physical systems (CPSs) by detecting potential cyber or physical disturbances. However, the development of scalable and optimal response measures under power system contingency based on fusing cyber-physical data is still in an early stage. To address this research gap, this paper introduces a power system response engine based on reinforcement learning (RL) and role and interaction discovery (RID) techniques. RL-RID-GridResponder is designed to automatically detect the contingency and assist with the decision-making process to ensure optimal power system operation. The RL-RID-GridResponder learns via an RL-based structure and achieves enhanced scalability by integrating an RID module with reduced action and state spaces. The applicability of RL-RID-GridResponder in providing scalable and optimal responses for CPSs is demonstrated on power systems in the context of Denial of Service (DoS) attacks.  Moreover, simulations are conducted on a Volt-Var regulation problem using the augmented WSCC 9-bus and augmented IEEE 24-bus systems based on fused cyber and physical data sets. The results show that the proposed RL-RID-GridResponder can provide fast and accurate responses to ensure optimal power system operation under DoS and can extend to other system contingencies such as line outages and loss of loads.

\end{abstract}

\maketitle

\section{Introduction}\label{sec:intro}

The assurance of resilience for critical cyber-physical systems (CPSs) is a multifaceted and challenging problem. High-impact low-frequency events, such as large-scale cyber or physical disturbances, can pose significant threats to power system reliability. In August 2023, the wildfire in Maui caused thousands of people to lose their homes. Hawaiian Electric was under scrutiny
for not cutting off electricity and not having proactive remedial reactions~\cite{Energy_hawaii}. This is not the first time the public has put the spotlight on utilities' emergency response ability. Historical real-world examples, such as the Ukraine attacks, illustrate that cyber attacks can severely disturb the operation of power systems, whether in steady or transient states. Moreover, the power system industry has gained a heightened awareness of the role of cybersecurity in achieving reliable operation for power systems~\cite{mitre,Nerc,NISTSP180032DER,NISTIR7628r1}. The National Institute of Standards and Technology Interagency Report documents the importance of a \textit{Defense-in-Depth strategy} in mitigating risks associated with cyber attacks \cite{NISTIR7628r1}. The strategy highlights the critical role of effective control and response against various types of cyber threats, aiming at maintaining multiple layers of security measures to prevent unauthorized access or disruption.

Power system resilience relies on the secure operation of both cyber and physical components and their crucial interdependencies \cite{5589488}.  Cyber and physical attacks, such as denial-of-service (DoS), man (machine)-in-the-middle (MiTM), and false data injection (FDI) attacks, can potentially affect both the dynamic and transient states of the system, which can lead to compromised resiliency and stability \cite{9225126,9707300,huang2024toward}. Therefore, major enhancements have been made to power system security against these threats over the past decade. These enhancements include developing cyber-aware grid planning and monitoring methods \cite{Nerc}, as well as remedial schemes to maintain voltage magnitudes, redirect power flows, and limit the effects of disturbances \cite{6032699,NERCRAS}. Nevertheless, new threat challenges continue to evolve within cyber-physical energy systems. On the energy side, the surging adoption of renewables introduces new scalability challenges due to their heterogeneity and numerosity \cite{lai2017cyber,Nerc2017}. On the security side, emerging advanced 
intrusion techniques result in an increased variety of disturbances that can affect both cyber and physical components in CPSs, highlighting the need for designing scalable and optimal cyber-physical response approaches \cite{FERC,2023PECI,6032699}.

With the advancements in computing algorithms and hardware, learning-based techniques are gaining rising attention in optimizing the operation of CPSs, such as guaranteeing cyber and physical security for CPSs, accelerating the integration of renewables for large-scale power systems, managing plug-in electric vehicles, as well as contributing to the general decision-making processes in CPSs \cite{8295075,9272624}. Among various learning-based methods, RL-based approaches, including state-action-reward-state-action (SARSA), deep reinforcement learning (DRL), and Q-learning, are considered highly promising methods in enhancing the optimal operation of power systems \cite{glavic2017reinforcement, BernadicKujundzicPrimorac2023,zhao2022reinforcement}. In RL-based approaches, agents interact with the dynamic environment through trial-and-error to explore various actions and obtain rewards, gradually refining their strategies to achieve optimal objectives \cite{kaelbling1996reinforcement}. The environment in which an agent operates is typically modeled as the Markov decision process (MDP), composing a pair of states, actions, transition probabilities, and rewards \cite{BernadicKujundzicPrimorac2023}. By providing adaptive decision-making capabilities, RL-based methods can optimize control strategies in real time and enhance the overall efficiency and reliability of the power system. Subsequently, RL has been studied in various power system applications, such as load frequency control, voltage control, economic dispatch, stability enhancement, relay control, and security analysis \cite{glavic2017reinforcement}. For example, RL can contribute to providing efficient energy management system (EMS) solutions, including demand response, flexible energy storage, and the integration of renewable energy sources \cite{BernadicKujundzicPrimorac2023}. Ernst \emph{et al.} \cite{1266597} discuss the potential of applying RL in power system stability control, highlighting the benefits through online and offline learning case studies. The RL-driven agents observe the system states, take actions, and learn from the outcomes, gradually accumulating experience to improve control strategies in damping power system oscillations \cite{1266597}. In \cite{pan2021improving}, a policy-based RL algorithm is designed for adversarial training, aiming to increase the robustness of RL agents against attacks and avoid infeasible operational decisions. Therefore, RL-based methods are powerful tools in ensuring grid resiliency and providing optimal grid management towards cyber-physical secure power system operation.

In addition to keeping up with evolving threats, another major challenge in deploying learning-based methods comes from their scalability issue when applied to complicated CPSs \cite{8909814}. The numerous actions and states for power systems, especially during certain contingencies, can easily overwhelm the decision-making process. Besides, power system responses can have localized, hierarchical, or centralized aspects that vary depending on the involved stakeholders and assets, largely complicating the decision-making process \cite{restore}.  Therefore, at each level, it is essential to make accurate decisions in a scalable manner 
to ensure timely response and prevent further damage \cite{glavic2017reinforcement}. To address this, the role and interaction discovery (RID), first presented in \cite{ISAP2017}, can generate reduced case-specific action spaces, helping the response engine address the dimension challenge in both cyber and physical remediation actions~\cite{ISAP2017,hossain2020enabling}. The RID was designed to identify essential, critical, and redundant controllers using clustering and factorization techniques based on the controllability of a CPS. Specifically, the \emph{Essential Controllers} determine the minimal set of devices required to maintain system controllability, the \emph{Critical Controllers} are essential controllers that occur in every minimal-cut controllability set of the system, the \emph{Redundant Controllers} are the devices that reinforce the control capability of essential controllers and can be removed without affecting system controllability, and the \emph{Control Support Groups} contain devices that are highly coupled in terms of impact on the control objective and with each other. The applicability of RID has been shown in multiple power system applications, such as corrective line switching to mitigate geomagnetically induced currents saturated reactive power losses \cite{hossain2020enabling}, reducing power system constraint violations by leveraging the characterization of distributed flexible AC transmission system controllers and generators \cite{hossain2021strategy,hossain2017analytic}. Despite its proven effectiveness, the integration of RID in RL-based power system environments with fused cyber-physical data for optimal response has not yet been addressed.


To this end, we propose a novel RL and RID-based GridResponder
aimed at advancing the accurate and fast intrusion detection and response for large-scale cyber-physical power systems. This paper significantly expands our previous conference paper~\cite{ScoreTPEC} that first outlines the motivation of RL-RID-GridResponder. Compared to \cite{ScoreTPEC}, this paper 
1) tailors and integrates the RID into RL-RID-GridResponder to enhance scalability with reduced action and state spaces; 2) fuses cyber and physical data to facilitate a fast and accurate decision-making process for power systems during DoS disturbances; 3) illustrates design of the proposed RL-RID-GridResponder in detailed submodules; 4) presents key experimental results and insights by implementing RL-RID-GridResponder on a cyber-physical power system testbed. After building the RL-based optimal response engine, we first construct a cyber-physical synthetic power system environment and then test it via real-time interaction at the testbed. The contributions are summarized as follows: 
\begin{itemize}
\item \textit{Data fusion based optimal response:} We fuse data that contain both cyber and physical features to assist optimal responses during power system contingency, i.e., DoS attack. 

\item \textit{Scalability:} 
The GridResponder can provide scalable RL-based solutions in real time for complicated cyber-physical power systems with extensive action and state spaces.

\item \textit{Response generalizability:} 
The proposed optimal response engine 
aids the system to operate robustly and resiliently under cyber or physical disturbances, and it is compatible with an extension on state-of-the-art RL-based structures. 

\item \textit{Optimality reassurance:} The designed response method serves as an optimal control strategy for managing various grid-tied resources. 

\end{itemize}


The rest of this paper is organized as follows: Section \ref{sec:background} introduces the preliminaries. In Section \ref{sec:submodules}, we detail the design of the RL-based scalable optimal response engine. Section \ref{sec:pr} provides experiments and analyses on a cyber-physical power system testbed. Section \ref{sec:conclusion} concludes the paper and provides future research directions.

\section{Preliminaries}\label{sec:background}

This paper introduces a scalable optimal response engine that can provide rapid and optimal responses during a contingency for cyber-physical power systems. This section presents related preliminaries.

\subsection{Data Security in CPS Control}\label{sub:datasecurity}

Data security threats from internal, external, and third-party sources can hinder the deployment of learning-based approaches for controlling critical CPSs
\cite{9116436}. The development of learning-based methods for operating critical CPSs, particularly in cyber
detection and response, must be done in a way that can be accurately validated, with verified safety, and quantified benefits for the application.
To this end, several standard metrics are commonly used to evaluate learning results, such as precision $\xi$, recall $\rho$, and $F_1$ score \cite{fscore}, where $\xi$ indicates the proportion of true positives (TP) divided by the total number of elements labeled positive, and $\rho$ is defined as the number of TP divided by the total number of actual positives. The $F_1$ score is defined as the harmonic mean of $\xi$ and $\rho$ to measure the test accuracy. Specifically, they are defined as   
\begin{subequations}\label{eq:PRF}
\begin{align}
\xi &= \frac{\text{TP}}{\text{TP} + \text{FP}},\\
\rho &= \frac{\text{TP}}{\text{TP} + \text{FN}},\\
F_1 &= \frac{2\xi \rho}{\xi+\rho},
\end{align}
\end{subequations}


\noindent where FP denotes false positives, i.e., the number of anomalies detected incorrectly, FN denotes false negatives, i.e., the number of anomalies that were missed in the detection. Therefore, an intrusion detection problem aims to miss fewer attacks by having high recall and low false alarms. These metrics can give a fair evaluation of the detection capabilities of learning-based methods, indicating the training and testing efficiency.

\subsection{Threat Model}\label{sec:threatmodel} 


The developed 
RL-RID-GridResponder aims to provide optimal cyber-physical responses for CPSs in the presence of both cyber and physical disturbances. In the following study,
we take the DoS attack for example
to demonstrate the response performance of the developed response engine. In a DoS attack, an adversary can disrupt or completely shut down a vital service or process, such as by flooding the target system with overwhelming traffic.  Further, each DoS attack has certain precursor steps that, once taken, may give an adversary even greater capabilities. These early-stage behaviors must also be monitored and learned to predict and prevent further disruptions. Therefore, the GridResponder employs RL to assist with cyber and physical controls at all stages of the data flow pipeline and to inform a response that is safe, fast, accurate, and interactive.

Note that GridResponder is not limited to protecting the CPS from DoS attacks but is rather designed for a generic threat landscape. For example, 
scenarios can occur in power systems when an intruder accesses communications between an operator and a critical device, such as a relay controller. The intruder can possibly alter commands sent by the operator to shut down or misoperate a relay, leading to severe consequences such as unexpected device behavior and loss of load, which could contribute to a cascading failure.  Essentially, the developed method would help understand how a learning-based scalable,  optimal, and real-time response engine can be attained during severe cyber or physical disturbances. The RL-based GridResponder is expected to act as a foundation for building real-world optimal response engines for CPSs, especially for power systems that are intrinsically complicated through their vast numbers of components.

\subsection{Testbed Emulation}\label{sub:ta}

Cyber-physical emulation is a crucial step for testing the efficiency and efficacy of a response engine. Due to the lack of a complete end-to-end control loop that is available out in the field for testing cyber-physical power systems, the deployment of existing optimal response techniques has largely been restrained \cite{anwar2017intrusion}. To remove this obstacle, this work performs testbed emulation by synthesizing and verifying the proposed approach within the \emph{Resilient Energy Systems Lab (RESLab)} testbed \cite{RESLab_our}.

The RESLab testbed can simulate the cyber-physical environment of realistic large-scale power systems and validate the response performance. The testbed consists of a power system interactive simulator that runs in near-real-time, a connected network emulator, intrusion detection systems like Snort \cite{snort_web}, Elastic Search-Kibana data aggregation and visualization \cite{Kibana_web}, hardware devices including protective relays and real-time automation control (RTACs), a proprietary platform from \emph{Schweitzer Engineering Laboratories (SEL)} \cite{SEL_web}, and a cyber-physical resilient energy system energy management system (CYPRES EMS) \cite{10024808}. The RESLab testbed has industrial control system protocols running through the emulated network and connecting with our power system simulator, our CYPRES EMS, and industrial hardware devices \cite{9521204}. 

Specifically, 1) \emph{PowerWorld Dynamic Studio (PWDS)} \cite{powerworld_web} serves as the real-time power system simulator that can be configured to send Distributed Network Protocol-3 (DNP3) outstation (OS) packets within realistic time constraints. With the DNP3 OS capability, it allows PWDS to act like a DNP3 server by sending data packages and communicating with other clients/servers \cite{huang2021real}; 2)  In RESLab, the \emph{Common Open Research Emulator (CORE)} is employed to simulate the communications between cyber components and other virtual machines (VMs) \cite{9521204}; 3) A GUI and console interface is applied in a software DNP3 master. Besides, an additional DNP3 master in a physical device, named \emph{SEL-3530 Real-time Automation Controller (RTAC)}, is configured to monitor and operate physical equipment in outstations; 4) The input/output signals of these actual protective devices are implemented in the monitor and control loop at the RESLab \cite{9808163}. They connect with the simulation in PWDS and emulation in CORE, respectively; 5) Snort serves as an intrusion detection engine that can be configured to detect DoS, MiTM, and address resolution protocol cache poisoning-based attacks and sends alerts to the DNP3 master; 6) The CYPRES EMS is designed to help analyze power systems from a security-oriented engineering perspective. The CYPRES EMS is an end-to-end system that performs cyber and physical network visualization, system monitoring and control, and mitigation.

To summarize, the RESLab testbed collects, stores, analyzes, and provides visualization of various power system information and is used in this paper to validate the RL-RID-GridResponder's performance.
















\section{Design of the RL-based Scalable
Optimal Response Engine}\label{sec:submodules}

\begin{figure*}[!hbt]%
  \centering
  \includegraphics[scale=0.60]{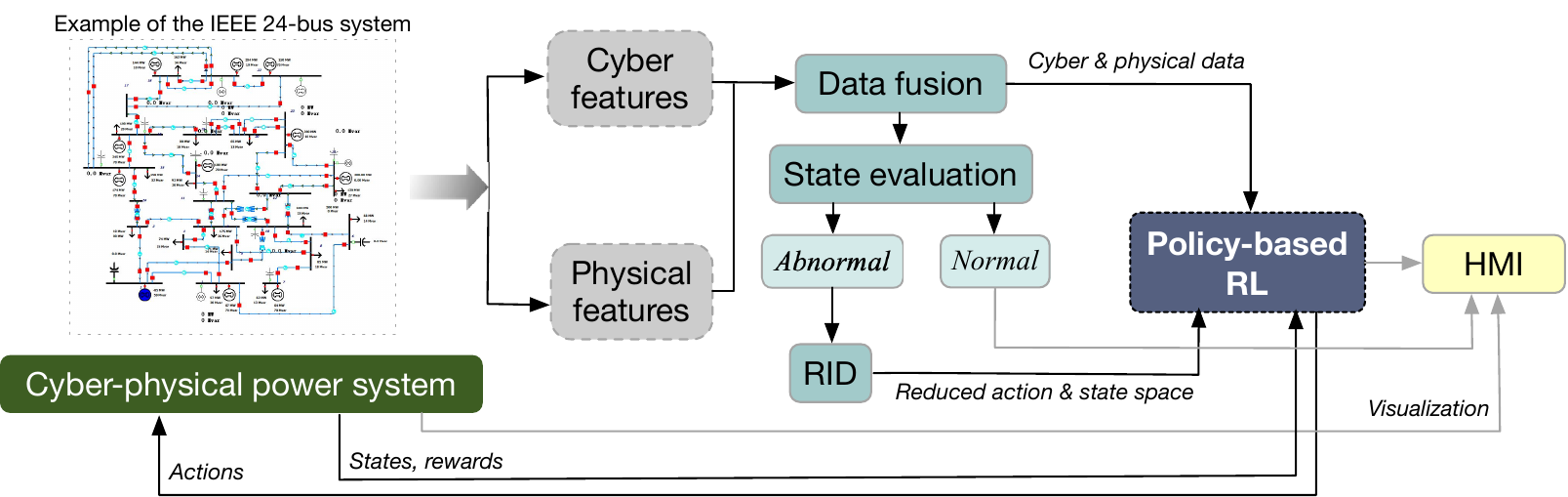}
 \caption{Framework of the proposed RL-based scalable optimal response engine for cyber-physical power systems.}
  \label{fig:rl}
\end{figure*}

This section details RL-RID-GridResponder's submodules, including the data fusion module, state evaluation module, RID module, RL module, and HMI module. We show their independent design and interoperation as an optimal response engine to help power systems return to an operationally secure and optimal state under a system contingency.  The generic framework of the proposed RL-RID-GridResponder is shown in Fig. \ref{fig:rl}.

\subsection{Data Fusion Module}
\label{section_data_fusion}

Cyber-physical data fusion refers to the process of integrating and analyzing data from various cyber and physical sources to ensure the system's optimality and integrity. It enhances the security of CPSs by effectively detecting anomalies, identifying potential cyber intrusions, and mitigating potential risks, combining information from different sources, such as sensors, control systems, and network logs. The data fusion module is initiated to assist RL-RID-GridResponder to identify an abnormal system status. By collecting both physical data from the power system and cyber data from the DNP3 packets, RL-RID-GridResponder enables multimodal analysis for assessing the power system operation. The multimodal analysis is designed to improve the data processing efficiency while analyzing system behavior under cyber-physical contingencies. In power systems, the compression of data to lower dimensions enhances storage and computing efficiency while retaining essential information, including round-trip time (RTT) delays and data from communication protocol packets, as well as bus voltages and currents from sensors. RTT is defined as the time duration between sending a DNP3 request and receiving the DNP3 acknowledgment message.  For example, in the IEEE 24-bus system, we monitor the magnitude of the generator at Bus 22 and the status of the transformer between Bus 24 and Bus 3 using DNP3 read acknowledgment messages from a separate virtual machine that acts as a DNP3 Master. The time interval between this communication is recorded as RTT. Additionally, during a DoS attack, when an adversary uses a direct DNP3 command to take the generator and transformer offline, the RTT is also recorded.

Specifically, we utilize principal component analysis (PCA) for data preprocessing and subsequently apply t-distributed stochastic neighbor embedding (t-SNE) to reduce the dimensionality of our dataset~\cite{haque2024multimodal}. Applying PCA before t-SNE can greatly speed up the t-SNE algorithm that can be computationally intensive with high-dimensional data. PCA also aids in removing noise by discarding low-variance dimensions that may represent noise rather than meaningful information. Therefore, PCA helps reduce dimensionality, making it easier for t-SNE to handle high-dimensional data effectively. 





In a high-dimensional space $\mathcal{X}$, t-SNE assesses pairwise similarities between data points using a Gaussian kernel, where closer points exhibit higher similarity. For a set of high-dimensional data points $x_1, x_2, ..., x_N$, the conditional probability $p_{ji}$ is defined to be proportional to the similarity between any two points, e.g., $x_i$ and $x_j$. Likewise, t-SNE constructs probability distributions $q_{ji}$ based on pairwise similarities in a lower-dimensional space $\mathcal{Y}$. To align $p_{ji}$ and $q_{ji}$, t-SNE minimizes the Kullback-Leibler (KL) divergence between the probability distributions $\mathcal{X}$ and $\mathcal{Y}$. The KL divergence can be obtained as 
\begin{equation}
K = \sum_i \mathrm{KL}(p_{i}||q_{i})=\sum_i\sum_jp_{ji}log\frac{p_{ji}}{q_{ji}}.
\label{cost_func}
\end{equation}

Then, the KL divergence can be minimized using the gradient descent as
\begin{equation}
\frac{\partial K}{\partial y_i}=4\sum_j(p_{ji}-q_{ji})(y_j-y_i){(1+||y_i-y_j||_2^2)}^{-1}.
\label{cost_minimi}
\end{equation}

Additionally, the data fusion module integrates real-time risk alerts, such as the Snort alerts that can provide insights into network intrusions, and other alerts through the CYPRES EMS that can offer guidance on system operational health and anomalies. These alerts are used as inputs to the data fusion module in the RESLab testbed. Then, the RL module interacts with the data fusion module to access the real-time data through local area networks. Simultaneously, real-time data are presented in an HMI for enhanced visualization.

To illustrate the idea, we present the analysis of two selected contingencies on the IEEE 24-bus system \cite{IEEE_RTS, WSCC_web}. These contingencies have been selected by performing a comprehensive contingency analysis on the power system case, and then identifying the contingencies that produced the most severe voltage violations in the case. In this way, the most severe contingencies were selected in this paper. In the first case, Use Case 1 (UC1), the transformer between Bus 24 and Bus 3 is compromised due to a cyber or physical threat. In the second case, Use Case 2 (UC2), both the generator at Bus 22 and the transformer between Bus 24 and Bus 3 are compromised due to a cyber or physical threat. For both cases, we consider physical features including bus voltage magnitudes, bus voltage angles, and current magnitudes through all branches. 

Additionally, cyber features have been added to the dataset. In this paper, we select the round-trip time (RTT) as the cyber feature to measure the communication latency during a DoS attack. The RTT is defined as the time duration between reading and acknowledging responses from DNP3 packets.  The RTT affects t-SNE clustering by grouping states with similar RTT values, thereby increasing the accuracy for clustering unstable states from stable ones. Moreover, in cases where other physical data is compromised, RTT remains a reliable indicator of system stability, allowing for actual state inference. Therefore, the data fusion module comprehensively combines both physical and cyber (communication) aspects to  clearly explain the power system states. Future work will also investigate additional cyber features or leveraging alerts from Snort or other Intrusion detection systems.  The cyber and physical features studied for the IEEE 24-bus case are exemplified in Table\ref{tab:multimodal}.

\begin{table}[htb]
\caption{An example of cyber and physical features for the IEEE 24-bus case.}
\label{tab:multimodal}
\begin{tabular}{|lll|}
\hline
\multicolumn{3}{|c|}{\textbf{Cyber Features}}                                                                   \\ \hline
\multicolumn{1}{|l|}{\textbf{Feature Name}} & \multicolumn{1}{l|}{\textbf{Number of Features}} & \textbf{Total} \\ \hline
\multicolumn{1}{|l|}{Round-trip time (RTT)}       & \multicolumn{1}{l|}{DNP3 communications}          & 1              \\ \hline
\multicolumn{3}{|c|}{\textbf{Physical Features}}                                                                \\ \hline
\multicolumn{1}{|l|}{\textbf{Feature Name}} & \multicolumn{1}{l|}{\textbf{Number of Features}} & \textbf{Total} \\ \hline
\multicolumn{1}{|l|}{Voltage magnitude}     & \multicolumn{1}{l|}{24 buses}                    & 24             \\ \hline
\multicolumn{1}{|l|}{Voltage angle}         & \multicolumn{1}{l|}{24 buses}                    & 24             \\ \hline
\multicolumn{1}{|l|}{Current magnitude}     & \multicolumn{1}{l|}{38 branches}                 & 38             \\ \hline
\multicolumn{2}{|c|}{\textbf{Total Number of   Features}}                                      & 87    \\ \hline
\end{tabular}%
\end{table}

The data fusion module merges data from cyber and physical sources to create an accurate  representation of an environment, therefore benefiting RL agents by providing richer information during decision-making. When applying data fusion for RL in power systems, agents can access multiple data sources, such as bus voltages and relay statuses through a communication network. By combining these data modalities, data fusion improves the agent's understanding of the power system status, handling uncertainty, and responding to contingencies more effectively. Moreover, data fusion provides historical context to help agents learn better policies in sequential decision-making tasks. More detailed experimental results on the IEEE 24-bus test case are given in Section \ref{Simu_IEEE_24_bus_test_case}.

\subsection{State Evaluation Module}

Based on real-time cyber and physical data inputs, the state evaluation module can assess the overall operational status of the current power system and determine whether or not the system is facing a disturbance or a cyber threat. Once under an abnormal status, the state evaluation module first classifies the types of disturbances, i.e., cyber, physical, or cyber-physical, then evaluates the consequence of the disturbance. A comprehensive assessment of whether the disturbances will cause failures or abnormalities within the physical, cyber, or cyber-physical domains can provide system operators with a thorough perspective on preparing and performing response strategies. Therefore, the state evaluation module can promote more accurate and effective responses to system disruptions, enabling the deployment of targeted mitigation strategies to restore normal operations and prevent further damage. After the state evaluation process, the real-time data is then processed by the RID module.

\subsection{Role and Interaction Discovery Module}\label{sec:rid}

The dimensionality curse remains a major challenge for RL-based methods, and this is also true for 
GridResponder that has to deal with high action and state spaces. The complexity of the action and state spaces grows significantly when considering both the cyber and physical features in designing the response engine. To enhance the algorithm scalability, we next focus on reducing the number of agents' action and state spaces by integrating the RID module. Despite its shown effectiveness in several power system control applications \cite{hossain2017analytic,hossain2020enabling,hossain2021strategy}, 
the integration of RID for identifying cyber and physical corrective actions within RL-RID-GridResponder has not yet been tackled before. Therefore, in this paper, we synthesize RID into the design of RL-RID-GridResponder to resolve the scalability obstacle, subsequently helping RL-RID-GridResponder efficiently utilize fused cyber-physical data and directly inform its RL-based decisions with \emph{a-priori} remediation action characterization.



The RID algorithm can be formulated via a three-step process as \cite{ISAP2017,hossain2022harmonized}: 
\begin{enumerate}
    \item \emph{Obtaining sensitivity matrix:} A linearized relationship between control actions and the system's response to those actions is provided by the sensitivity matrix $\bf{\Psi}$.  In this paper, the sensitivity $\bf{\Psi}$ indicates the relations between the actual power injection from the generation sources and the real power flow $\Delta P$ on each line that is experiencing overload, and is defined as  
    \begin{equation}\label{eq:line}
        \Delta P_{\mathrm{flow.line,overloaded}} =[\bm{\Psi}]  \cdot \Delta G_\mathrm{MW}.
    \end{equation}
The sensitivities also help us understand the relationship between the capacity of available capacitors and buses under overload conditions, by illustrating how the voltage levels $\Delta V$ at each overloaded bus respond to variations in reactive power $\Delta Q$ supplied by newly integrated capacitors as
\begin{equation}\label{eq:bus}
    \Delta V_{\mathrm{bus,overloaded}} =[\bm{\Psi}]   \cdot \Delta Q_\mathrm{MVar}.
\end{equation}

Eqs. \eqref{eq:line} and \eqref{eq:bus} assist in comprehensively understanding the system's behavior under various control actions and enhance system's response ability.
     
\item \emph{Finding controllability-equivalence sets:} The control support groups are determined by clustering the sensitivity matrix rows that exhibit the mutual influence of controls within the different controllers. The similarity of row vectors $\mathbf{v}_i$ and $\mathbf{v}_j$ is determined by the coupling index (CI), which is the cosine similarity as follows: 
    
    \begin{equation}
\mathrm{CI} = \cos(\theta_{\mathbf{v}_i \mathbf{v}_j}) = \frac{\mathbf{v}_i \cdot \mathbf{v}_j}{\|\mathbf{v}_i\|\|\mathbf{v}_j\|}. 
\end{equation}

\item \emph{Finding critical, essential, and redundant sets:} The columns of $\bf{\Psi}$ are used to identify the critical, essential, and redundant controllers. This decomposition is important for allowing one to identify what different controller types are present in the system, where they are located, and for developing response and mitigation applications based on these roles. As detailed in \cite{ISAP2017,hossain2020enabling}, the RID performs a change of basis that maps available controllers to equivalent controllable states. 
From the LU factorization, a change of basis is performed to decompose the transposed sensitivity matrix to lower and upper triangular factors. The detailed explanation of the LU factorization can be found in \cite{Peters01011970}. The factorization of $\bf{[\Psi]}^\textsc{T}$  can be represented as follows:
\begin{equation}
[\bm{\Psi}]^\textsc{T}=\mathbf{P}^{-1}\mathbf{L}_{\mathrm{F}} \mathbf{U}_{\mathrm{F}},
\label{eq:lu}
\end{equation}
\begin{equation}
\mathbf{L}_{\mathrm{F}} = \begin{bmatrix}
           \mathbf{L}_{\mathrm{b}} \\
           \mathbf{M}
         \end{bmatrix}.
         \label{redun}
\end{equation}

Based on the Peters-Wilkinson method~\cite{Peters01011970}, the matrix $\mathbf{[\Psi]}^\textsc{T}$ is decomposed, with $\bf{P}$ representing the permutation matrix, and $ \mathbf{L}_{\mathrm{F}}$ and $\mathbf{U}_{\mathrm{F}}$ being the lower and upper triangular factors of the matrix dimension, respectively. Additionally, $\bf{M}$ is identified as a sparse, rectangular matrix. The new basis is obtained as:
\begin{equation}
   \mathbf{L}_{\mathrm{CER}} = \mathbf{L}_{\mathrm{F}} \mathbf{L}_{\mathrm{b}}^{-1} = \begin{bmatrix}
           \mathbf{C_{\mathrm{E}}} \\
           \mathbf{C_{\mathrm{R}}}
         \end{bmatrix},
         \label{basis}
\end{equation} 
with each row of the transformed matrix corresponding to its available controller~\cite{ChenAbur2006}. $\mathbf{C_{\mathrm{E}}}$ denotes the identity matrix $\mathbf{I}_n$, with rows corresponding to essential controllers. The rows of $\mathbf{C_{\mathrm{R}}}$ correspond to redundant controllers. Columns correspond to equivalent controlled states, e.g., overloaded line flows, which can be easily mapped back to the original flows using $\bf{P}$.


\end{enumerate}

\subsection{Reinforcement Learning Module}
Previous works include supporting the system with optimal solutions through the MDP process by optimizing it using value iteration and policy iteration approaches \cite{8909814}. Building on previous optimization approaches, RL offers a more dynamic and flexible solution.
RL's adaptability
in continuously learning and making decisions is a key characteristic that makes
RL-based paradigms 
 %
 well suited 
 for the design of an optimal cyber-physical response engine. 
 In RL, an agent learns to make optimal decisions by interacting with the environment and receiving feedback in the form of rewards or penalties \cite{kaelbling1996reinforcement}. In designing RL-RID-GridResponder, the possible agents in the model could be the physical components, like capacitors or generators, 
 or the cyber components, such as firewalls or routers, and the cyber-physical components, such as relays and remote terminal units.

RL can be formulated into a Markov decision process (MDP) where the agents learn to maximize the expected cumulative reward over time. Central to this framework are the concepts of states $s$, actions $a$, transition probabilities $P$, rewards $R$, and the discount factor $\gamma$ \cite{white1989markov, sutton1998reinforcement}. Specifically, the transition probabilities $P(s'|s,a)$, expected rewards $R(s,a)$, and value function $V(s)$ are defined as: 
\begin{subequations}
\label{eq_MDP}
\begin{align}
    P(s'|s,a) &= \Pr(s_{t+1} = s' | s_t = s, a_t = a),\\
    R(s,a) &= \mathbb{E}[r_{t+1} | s_t = s, a_t = a],\\
    V(s) &= \mathbb{E}[\sum_{k=0}^{\infty } \gamma ^kr_{t+k+1}|s_t=s],
\end{align}
\end{subequations} where $\gamma$ denotes the discount factor that prioritizes rewards of short-term rewards over long-term ones, $s$ denotes the current state, $s'$ denotes the future state, and $k$ is the execution time step. 





\subsubsection{Actions, States, and Rewards}

In what follows, we take the Volt-Var optimization problem as an example, to show the design of the GridResponder
By formulating the Volt-Var optimization problem into a MDP,  we have
\begin{itemize}
\vspace{-2mm}
    \item \textit{State space:} The state space consists of bus voltages, capacitor status, and tap changing transformer status.
    \item \textit{Action space:} The action space consists of both discrete and continuous actions. Discrete actions include the capacitor bank (on/off), voltage regulator tap number, and battery states in charge or discharge. Continuous actions include battery charging/discharging power in $[-1,1]$, where -1 represents fully charging and 1 represents fully discharging.
    \item \textit{Reward model:} The reward model is given by: 
    \begin{equation}
    R(s,{s}') = - F_{\mathrm{volt}}({s}') - F_{\mathrm{ctrl}}(s,{s}') - F_{\mathrm{power}}({s}'),
    \label{reward}
    \end{equation}
\end{itemize}
where $F_{\mathrm{volt}}(\cdot)$ denotes the voltage violation that is defined by:
\begin{equation}
    F_{\mathrm{volt}}({s}') = \sum_{n \in N}{\left ( V_n({s}') - \bar{V} \right )}_{+} + \sum_{n \in N}{\left ( \underbar{$V$} - V_n({s}')\right )}_{+},
    \label{f_volt}
\end{equation}
where $_{+}$ is a shorthand for $\max(\cdot, 0)$, $V_n$ is the voltage of bus $n$, $\bar{V}$ and $\underbar{$V$}$  are the upper and lower voltage limits, respectively.

$F_{\mathrm{ctrl}}(\cdot)$ denotes the sum of control errors from capacitors, regulating transformers, and batteries, and is given by:            
\begin{equation}
    \begin{aligned}
        F_{\mathrm{ctrl}}(s, {s}') = &\sum_{c \in \mathbb{C}}W_{\mathrm{cap}}\left | \mathrm{Status}_c(s) - \mathrm{Status}_c ({s}')\right | \\  &+  
    \sum_{r \in \mathbb{G}} W_{\mathrm{reg}}\left | \mathrm{Tap}_r(s) - \mathrm{Tap}_r ({s}')\right | \\&+
    \sum_{b \in \mathbb{B}}W_{\mathrm{dis}}\frac{D_b({s}')}{\bar D_b} + W_{\text{SoC}}\left  | \mathrm{SoC}_\mathrm{b}(s) - \text{SoC}_\mathrm{b}0\right |,
   \end{aligned}
\label{f_ctrl}
\end{equation} 
where subscripts $c$, $r$, and $b$ denote a capacitor, a regulating transformer, and a battery, respectively, $W_{\mathrm{cap}}$, $W_{\mathrm{reg}}$, $W_{\mathrm{dis}}$ and $W_{\mathrm{SoC}}$ $\in [0,1]$ denote the weights for controlling the capacitor, the regulating transformer, charging/discharging power of the battery, and battery SoC, respectively. $\mathrm{Status}_c (\cdot)$ equals to 1 when a capacitor is connected to the bus, and 0 when disconnected, $\mathrm{Tap}_r (\cdot)$ denotes the tap number of the regulating transformer, $D_b$ denotes discharge power and $\bar{D}_b$ denotes the max power, $\text{SoC}_\mathrm{b}(s) \in [0,1]$ and $\text{SoC}_\mathrm{b}0$ denote the current SoC of the battery and the initial SoC of the battery, respectively. This objective function penalizes the agent for frequently altering the status of a capacitor, the tap number of a transformer, or the SoC of a battery. As a result, an equilibrium is achieved with the minimum number of adjustments.

$F_{\mathrm{power}}(\cdot)$ denotes the power loss objective that is a ratio of the overall power loss to the total power, and is formulated as:
\begin{equation}
    F_{\mathrm{power}}({s}') = W_{\mathrm{power}}\frac{P_{\mathrm{loss}}({s}')}{P_{\mathrm{total}}({s}')},
    \label{f_power}
\end{equation}
where $W_{\mathrm{power}} \in [0,1]$ denotes the control weight.

\subsubsection{PPO and A2C}

Within the realm of RL, a variety of algorithms, including SARSA, deep RL (DRL), and Q-learning, have been proposed. Among them, DRL can handle complex and high-dimensional action and state spaces by combining RL with deep neural networks \cite{9275593}. A wide range of DRL architectures have been designed for training agents to make decisions in complex cyber and physical environments, including proximal policy optimization (PPO) \cite{schulman2017proximal}, advantage actor-critic (A2C) \cite{mnih2016asynchronous}, and deep Q-leaning \cite{fan2020theoretical}. In this paper, we illustrate the design of RL-based GridResponder via both PPO and A2C and provide analyses of cyber-physical power system applications.

PPO is an online policy-based RL algorithm that focuses on optimizing a parameterized policy using policy gradient \cite{PPO}.  It compares the new policy $\pi_{\theta}$ against the old policy $\pi_{\theta_{\mathrm{old}}}$ by computing the following probability ratio $r_t(\theta)$ as:

\begin{equation}
r_t(\theta) = \frac{\pi_{\theta}(a_t | s_t)}{\pi_{\theta_{\mathrm{old}}}(a_t | s_t)}.
\end{equation}

PPO can achieve the optimal policy by introducing a clipped surrogate objective function as:
 \begin{equation}
L^{\mathrm{clip}}(\theta) = \hat{\mathbb{E}}_t \left[ \min( r_t(\theta) \hat{A}_t, \mathrm{clip}(r_t(\theta), 1 - \epsilon, 1 + \epsilon) \hat{A}_t ) \right],
\end{equation} 
where $\epsilon$ denotes a hyperparameter that controls the clipping range, $\mathrm{clip}(\cdot)$ is the clipping function, and $\hat{A}_t$ is the advantage estimate at time $t$. The clipping operation prevents the probability ratio $r_t(\theta)$ from straying too far from the old policy, i.e., if $r_t(\theta)$ is within the range $[1 - \epsilon, 1 + \epsilon]$, it remains unchanged, otherwise, it is clipped back. The term $r_t(\theta) A_t$ represents the standard policy gradient objective. By taking the minimum of the clipped and non-clipped probability ratios, the objective function maintains a conservative update.

For A2C, it combines policy gradient methods (actor) and value function methods (critic) by adopting a policy function $\pi (a_t|s_t; \theta)$ and a value function $V(s_t; \omega)$. Specifically, the actor learns the policy $\pi (a_t|s_t; \theta)$, parameterized by $\theta$, that maps states $s_t$ to a probability distribution over actions $a_t$. The critic learns the value function $V(s_t; \omega)$, parameterized by $\omega$, that estimates the expected return (cumulative future rewards) from state $s_t$.

The advantage function  $A(s_t,a_t)$ is defined as:
\begin{equation}
    A(s_t,a_t)=Q(s_t,a_t) - V(s_t),
    \label{eq:a2c_1}
\end{equation}
where $Q(s_t,a_t)$ denotes the action-value function and  $V(s_t)$ is the state-value function. In practice, the advantage function can be approximated using the Temporal Difference (TD) error:
\begin{equation}
    A(s_t,a_t) \approx \delta_t = r_t + \gamma V(s_{t+1};\omega) - V(s_t;\omega),
    \label{eq:a2c_2}
\end{equation}
where $\delta_t$ denotes the TD error at time
step $t$, $\gamma$ denotes the discount factor, and $V(s_t;\omega)$ is the estimated value of state $s_t$. The policy parameters $\theta$ are updated using the policy gradient:
\begin{equation}
    \theta =\theta +\alpha \sum_{t}\nabla_\theta log \pi (a_t|s_t;\theta)\delta_t,
    \label{eq:a2c_3}
\end{equation}
where $\alpha$ denotes the learning rate for the actor. The value function parameter $w$ is updated by minimizing the squared TD error:

\begin{equation}
    \omega = \omega -\beta \sum_{t} \nabla_{\omega} (\delta_t)^2,
    \label{a2c_4}
\end{equation}
where $\beta$ denotes the learning rate for the critic.


The pseudo-code of the proposed RL-based GridResponder is given by Algorithm \ref{score}.

\RestyleAlgo{ruled} 

\begin{algorithm}
\caption{Pseudo code of the RL-RID-GridResponder}\label{score}
 Input: Real-time data through vSphere.\\
 
Data fusion with multimodal analysis, minimization of the KL divergence using \eqref{cost_func} and \eqref{cost_minimi};\\

State evaluation;\\

\If{State is \textit{abnormal}}{
Send abnormal alert to RL-RID-GridResponder;\\
Apply RID control using Eqs. \eqref{eq:line}-\eqref{basis};\\
Reduce agents' action and state spaces based on RID for the RL module;

\For {Training RL model}{
Initialize agents, action and state spaces, and rewards;\\
Initialize the cyber-physical environment;\\
Train RL agents by PPO or A2C\\
    $L^{\mathrm{clip}}(\theta) = \hat{\mathbb{E}}_t \left[ \min( r_t(\theta) \hat{A}_t, \mathrm{clip}(r_t(\theta), 1 - \epsilon, 1 + \epsilon) \hat{A}_t ) \right]$ or 
    $ A(s_t,a_t) \approx \delta_t =r_t + \gamma V(s_{t+1};\omega) - V(s_t;\omega)$ and
    $\theta =\theta +\alpha \sum_{t}\nabla_\theta log \pi (a_t|s_t;\theta)\delta_t$ and\\
    $\omega = \omega -\beta \sum_{t} \nabla_{\omega} (\delta_t)^2$;\\
 Compute rewards and update hyper-parameters;}
}
Visualisation: Via HMI.
\end{algorithm}


    
    
    

\subsection{Human Machine Interface Module}

The Human Machine Interface (HMI) plays an essential role in the
visualizations of cyber-physical state information and the 
user interactions with the analyses.
After evaluation and contingency analysis, the optimal response commands recommended by the RL-RID-GridResponder system will be sent to an HMI, where
the HMI's functions include displaying final commands, enabling user interaction, and collecting real-time feedback. 
Instead of relying solely on machine support, 
an ongoing direction for RL-RID-GridResponder is to integrate expertise of operators and analysts using feedback in the RL via the HMI to develop more dynamic solutions. For example, if the subject matter expert using the engine finds a recommendation wrong or unreasonable, they can provide negative feedback and apply their own solutions. In return, these user-provided rewards or feedback can further refine the RL model.

As shown in Fig. \ref{fig:rl}, 
RL-RID-GridResponder includes a data fusion module, a state evaluation module, and an RID module, with 
data pre-processing and analyzing procedures.  
After identifying a contingency, the proposed RL-RID-GridResponder provides real-time responses to mitigate the contingency and optimize the power system operation.
This approach is applicable to both distribution and transmission systems because the same techniques can be used with either kind of system.  Hence,
the approach has been developed without loss of generality.
Finally, the HMI module further offers visualization and calibration for the optimal response results. 

\section{Experimental Results} \label{sec:pr}

In this section, we present the experimental results of applying the developed RL-RID-GridResponder to assist with restoring the system to a steady and optimal state under a DoS attack. By "optimal," we mean that the RL process aims to achieve the minimum possible loss in terms of system performance. The MDP process is scaled over 24 hours with an hourly control frequency. Two policy-based RL algorithms, 
i.e., PPO and A2C, are integrated into GridResponder to benchmark the optimal response results. The objective is to optimize the voltage of each bus, such that the per-unit voltages remain within the limit of $[0.95,1.05]$ \cite{ANSI}. We take capacitors, batteries, and the tap of the transformers as RL agents to regulate the bus voltages. However, during a DoS attack, the system's ability to communicate with the battery is compromised, preventing the battery from assisting in adjusting the system states. Two test cases are conducted in the RESlab testbed for the Volt-Var control problem on the WSCC 9-bus system and on the IEEE 24-bus system, respectively.

\subsection{Simulation Environment}

Various open-source RL environments have been developed for large-scale power systems, such as Grid2op \cite{grid2op}, PowerGym \cite{fan2022powergym}, and the cyber-resilient power distribution environment \cite{sahu2023reinforcement}. These environments, built on top of OpenAI Gym, can serve as benchmarks for training RL algorithms. PowerGym, developed by Siemens, particularly focuses on Volt-Var control for distribution systems, utilizing OpenDSS as a back-end power flow solver to facilitate state and reward updates.



Therefore, in this paper, the RL environment is constructed based on an augmented version of the PowerGym framework. We enhanced the PowerGym framework to integrate the augmented WSCC 9-bus system and the IEEE 24-bus system. Besides, these systems have been meticulously upgraded to ensure compatibility with OpenDSS. The developed environment interacts with PowerWorld Simulator which is used as the backend in the RESLab testbed. Moreover, the environment includes a cyber layer implemented in Python that simulates a virtual network. The virtual network is designed to facilitate communication between the system components using the DNP3 communication protocol. This integration allows for realistic simulation and testing of CPSs, providing a robust platform for the development and evaluation of RL algorithms in power system applications. Therefore, this environment contains both cyber, physical, and cyber-physical components of the power system. Power flow analysis and contingency analysis are also  incorporated into the environment to  evaluate the performance of RL agents in real-time.



\subsection{Volt-Var Control}\label{sub:volt-var}

This paper presents two case studies of the Volt-Var control problem on the WSCC 9-bus and the IEEE 24-bus systems, respectively. In the WSCC 9-bus system, three available actions, including the real power source (battery bank), the reactive power source (capacitor banks), and tap changing transformer, can be controlled to optimize the voltage. Additionally, two regular synchronous machine generators are assumed to be in different control systems. This is realistic, as the generators are commonly controlled by a generator management system and rely on SCADA communications with the utility. We also make this experiment assumption for the IEEE 24-bus test case. Therefore, the controllable assets can provide response measures when the system is facing DoS attacks.

\subsubsection{WSCC 9-bus test case}

\begin{figure}[!hbt]%
  \centering
  \includegraphics[scale=0.25]{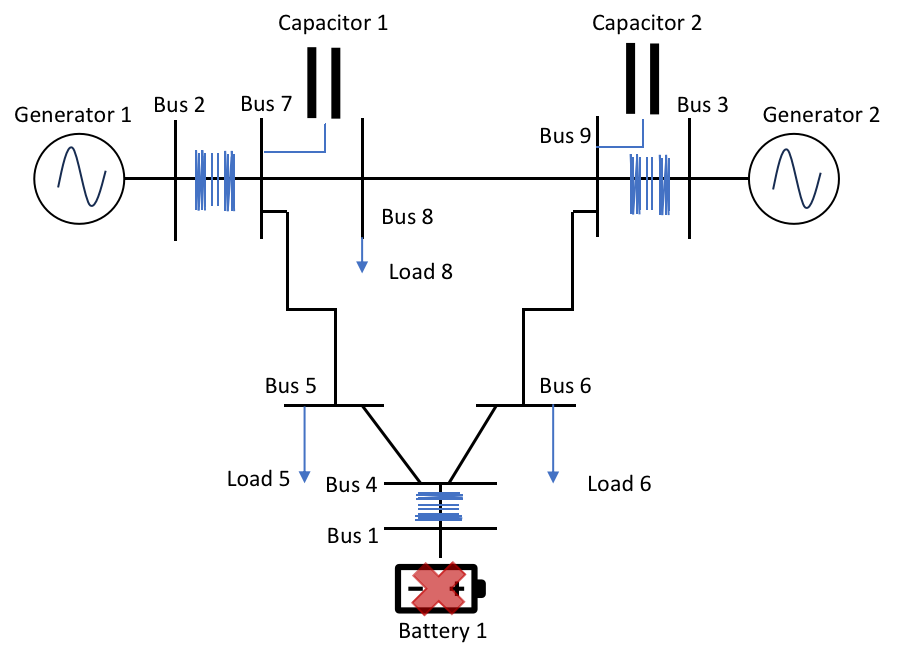}
 \caption{Augmented WSCC 9-bus system.
 }
  \label{fig:9bus}
\end{figure}

The augmented WSCC 9-bus system is shown in Fig. \ref{fig:9bus}.  

The WSCC 9-bus system has nine buses, three generators, and three substations with 21 network nodes in its synthetic cyber-physical model \cite{WSCC_web,al2000voltage}. We extracted the WSCC 9-bus model from PowerWorld Simulator and modified it to run in OpenDSS, which is compatible with the PowerGym environment. Additionally, the Generator 3 is replaced by a battery source. As shown in Fig. \ref{fig:9bus}, two 4000 kVAR capacitors are added to Bus 7 and Bus 9, respectively. In this case, transformers located at Substations A, B, and C are treated as online tap-changing transformers (voltage regulators). Other parameters are identical with \cite{al2000voltage}.

\begin{figure}
     \begin{subfigure}[b]{0.43\textwidth}
         \centering
         \includegraphics[width=\textwidth]{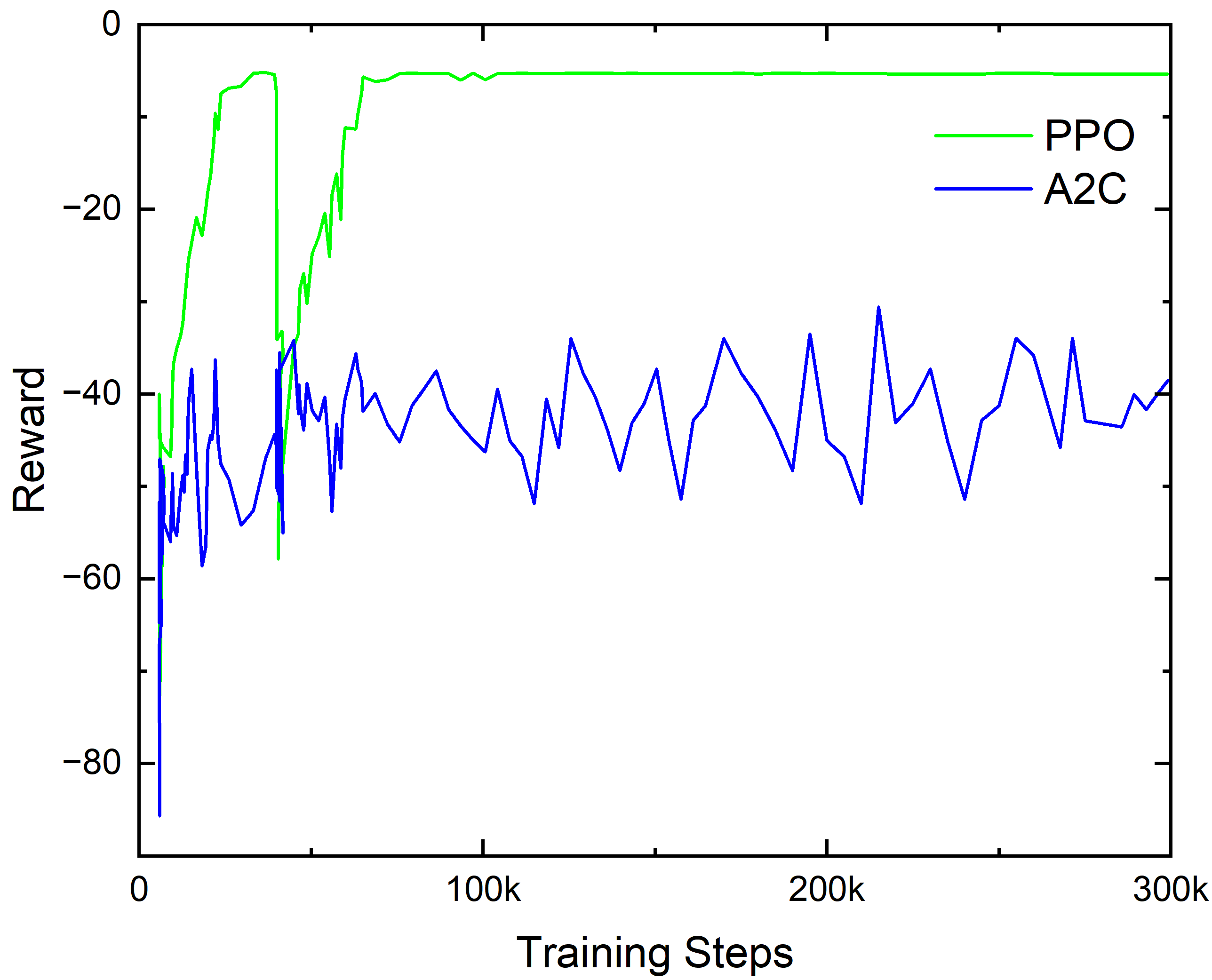}
         \caption{Rewards in RL-RID-GridResponder 
        under normal conditions.}
         \label{fig:reward}
     \end{subfigure}
     \hfill
     \vspace{4mm}
     \begin{subfigure}[b]{0.43\textwidth}
        \centering
         \includegraphics[width=\textwidth]{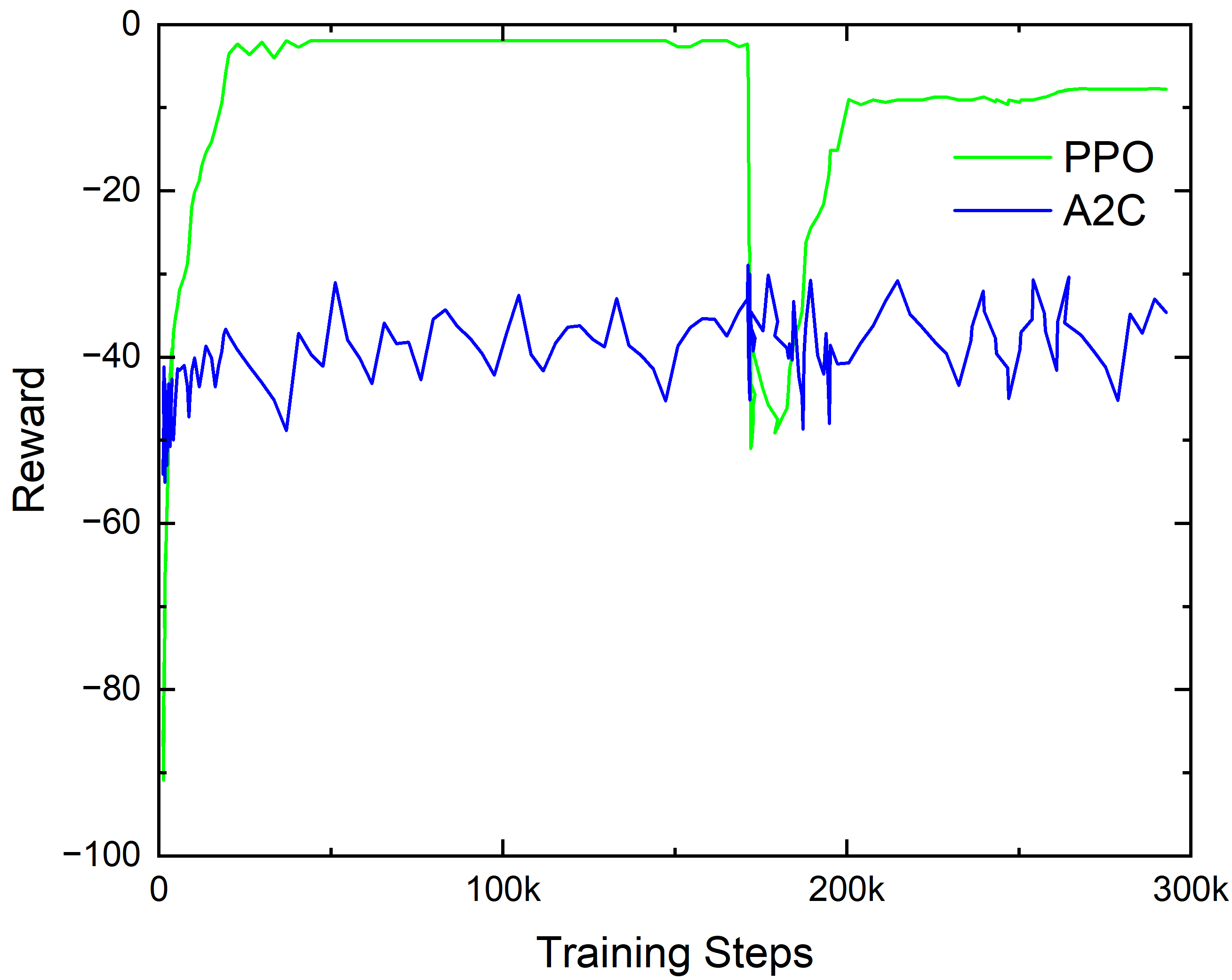}
         \caption{Rewards in RL-RID-GridResponder under DoS on Battery 1.}
         \label{fig:reward2}
     \end{subfigure}
        \caption{Comparison of RL results for PPO and A2C with/without the DoS, showing that both PPO and A2C converged while the PPO outperformed the A2C within fewer oscillations and better rewards.}
        \label{fig:rl9}
\end{figure}

We assume that Battery 1 faces a DoS attack. The RID algorithm is applied to the WSCC-9 bus system after the attack, with Table~\ref{tab:9bus} referring to the critical controllers, including Capacitors 1 and 2. To verify the effectiveness of the proposed RL-based GridResponder, we obtain the experimental results under two policy-based RL algorithms, i.e., PPO and A2C.  The results of normal state rewards and rewards facing cyber-disturbance are shown in Fig. \ref{fig:reward} and Fig. \ref{fig:reward2}, respectively.

\begin{table}[h]

\caption{RID result for the augmented WSCC 9-bus system.}
\begin{tabular}{|l|l|}
\hline
\textbf{Violations} & \textbf{Critical Controllers} \\ \hline
Bus voltage violations & Capacitors 1, 2 \\ \hline
\end{tabular}%
\label{tab:9bus}
\end{table} 

The load profiles are normalized to be within the range of $[0,1]$ and capture realistic load fluctuations including where under heavy load the voltages decrease and periods of light load where voltages rise. 
For every episode in 24 hours, three random loads are initialized hourly for each bus, and the reward is averaged.  

\begin{figure}[bt]
\centerline{\includegraphics[scale=0.70]{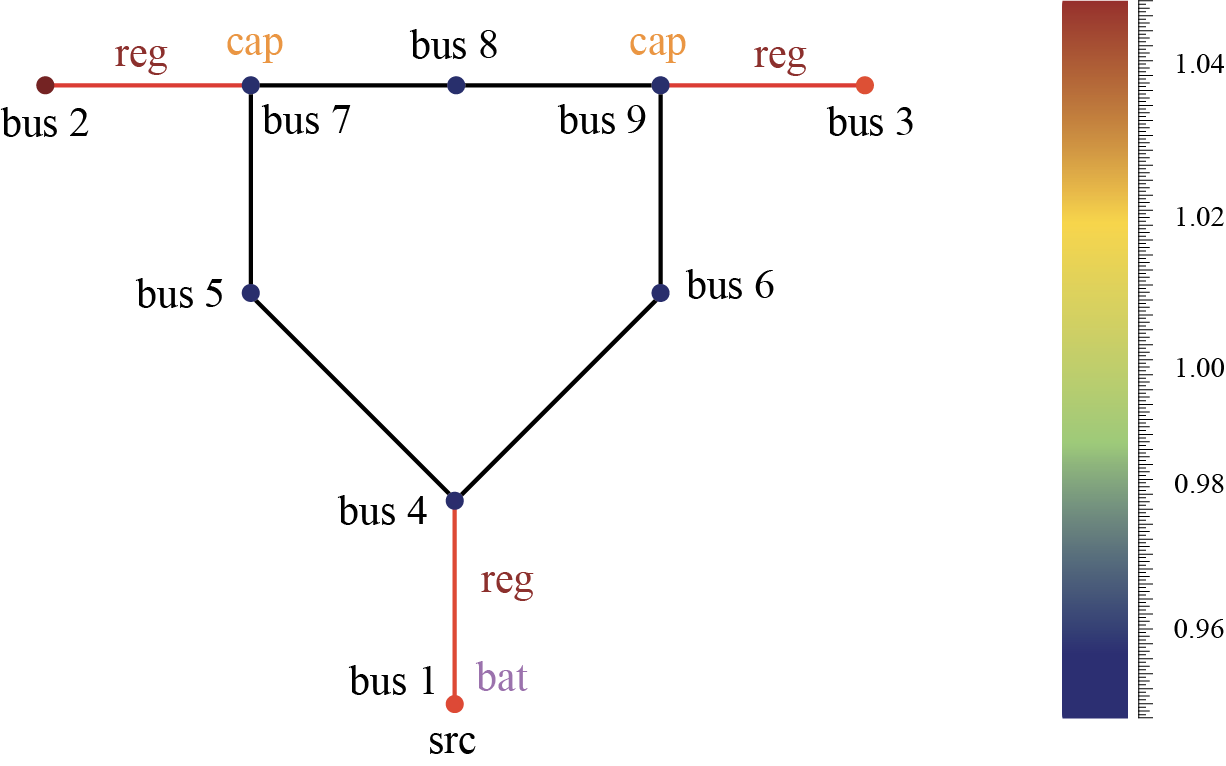}}
\caption{Bus voltages of the augmented WSCC 9-bus system presented via a heatmap (trained with PPO, where `cap', `bat', and `reg' denote capacitor, battery, and the tap of the transformer, respectively). All buses' voltages are kept within the ±5\% fluctuation.}
\label{fig:RL validation}
\end{figure}

The results show that after a sufficient amount of training, the reward settles down at a lower stable value. Observing from Fig. \ref{fig:reward} and Fig. \ref{fig:reward2}, it can be seen that the optimal reward of PPO is significantly higher than A2C. In both scenarios, the converged result of PPO is approximately 0, whereas A2C typically converges to values ranging from -40 to -50. When under DoS disturbance, the rewards of PPO are also significantly lower than A2C. Therefore, the PPO tends to outperform the A2C with fewer steps and with better rewards in this scenario. Additionally, the Volt-Var regulation results are shown in Fig. \ref{fig:RL validation}, where the voltages of all buses were kept within the $\pm 5\%$ fluctuation.

\subsubsection{IEEE 24-bus test case}
\label{Simu_IEEE_24_bus_test_case}

\begin{figure}[h]%
  \centering
  \includegraphics[scale=0.16]{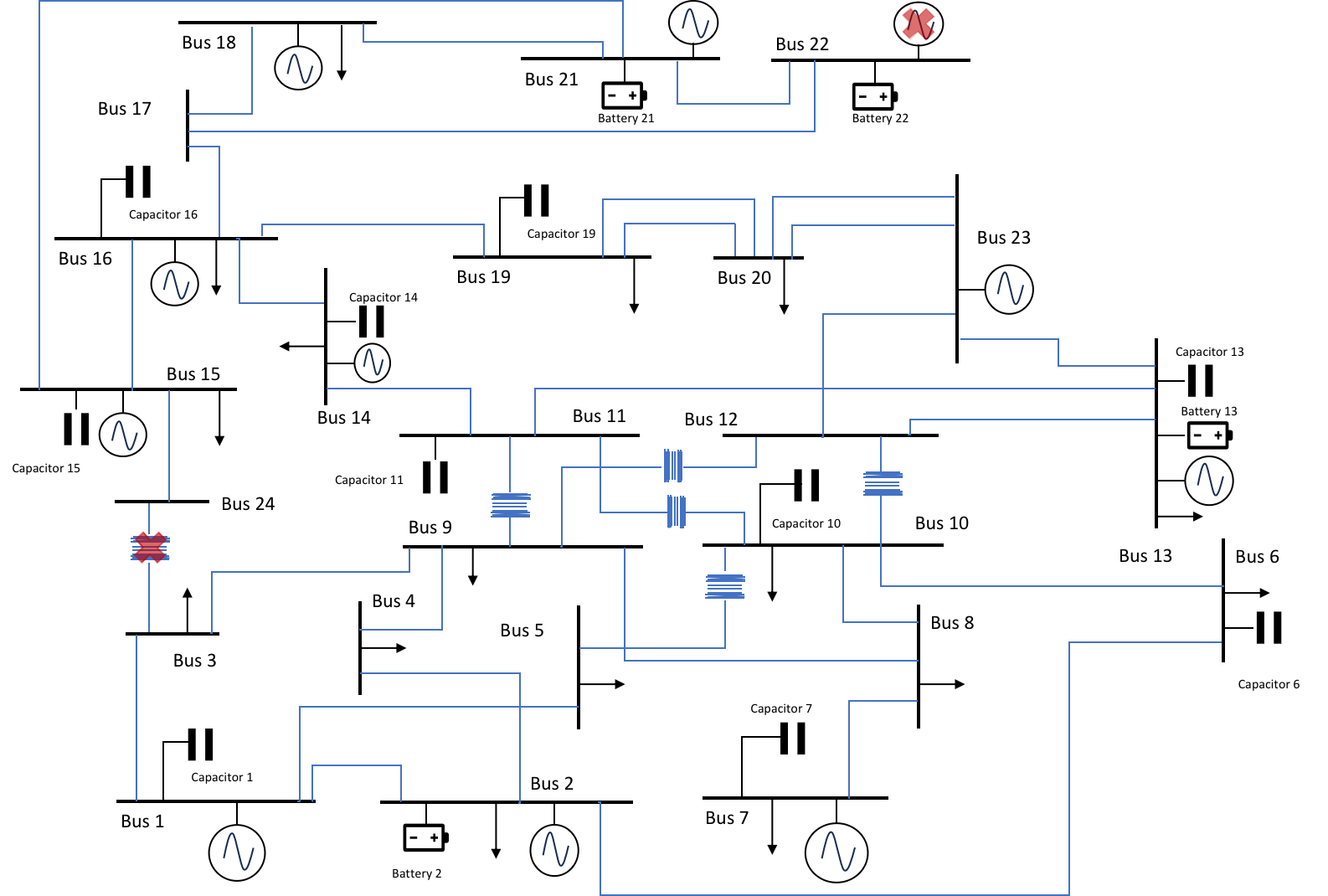}
 \caption{Augmented IEEE 24-bus system (with additional capacitors and batteries).}
  \label{fig:24bus}
\end{figure}

An IEEE 24-bus test system is utilized as another test case \cite{IEEE_RTS, WSCC_web}. The IEEE 24-bus test system includes 11 generators, six loads, and a single substation system with two networking nodes. Additionally, four batteries and nine capacitor banks are augmented when establishing the RL environment, i.e., nine capacitors are added on buses 6, 7, 10, 11, 11, 13, 14, 15, 16, 19, and four batteries are added to buses 2, 6, 21, 22. The augmented IEEE 24-bus system is shown in Fig.~\ref{fig:24bus}. 

The experiments are conducted
under two DoS use cases, including UC1 and UC2 as described in Section \ref{section_data_fusion}. These two use cases are selected based on a contingency analysis performed on the augmented IEEE 24-bus system using PowerWorld \cite{powerworld_web}, because these outages can result in primary voltage violations. By applying the RID algorithm, the essential and critical controllers are identified in Table \ref{tab:RID_24bus} for both use cases. Notably, in UC2, the outage of Transformer 24 and Generator 22 causes 
voltage instability and overloads on other transformers and generators.
As in Table~\ref{tab:RID_24bus}, under UC1, both Battery 2 and Battery 13 are identified as essential controllers, but only Battery 2 is considered critical, indicating that Battery 2 has a more significant role in managing outages in this scenario. Under UC2, a combination of four capacitors (Capacitors 11, 6, 10, 15) and two batteries (Batteries 2 and 13) are categorized as essential. Note that only Capacitors 6 and 10 are classified as critical in this scenario, and notably, there are no critical batteries in this scenario. 


\begin{figure*}[!h]
    \centering
    \begin{subfigure}{0.45\textwidth}
         \centering
         \includegraphics[width=\textwidth]{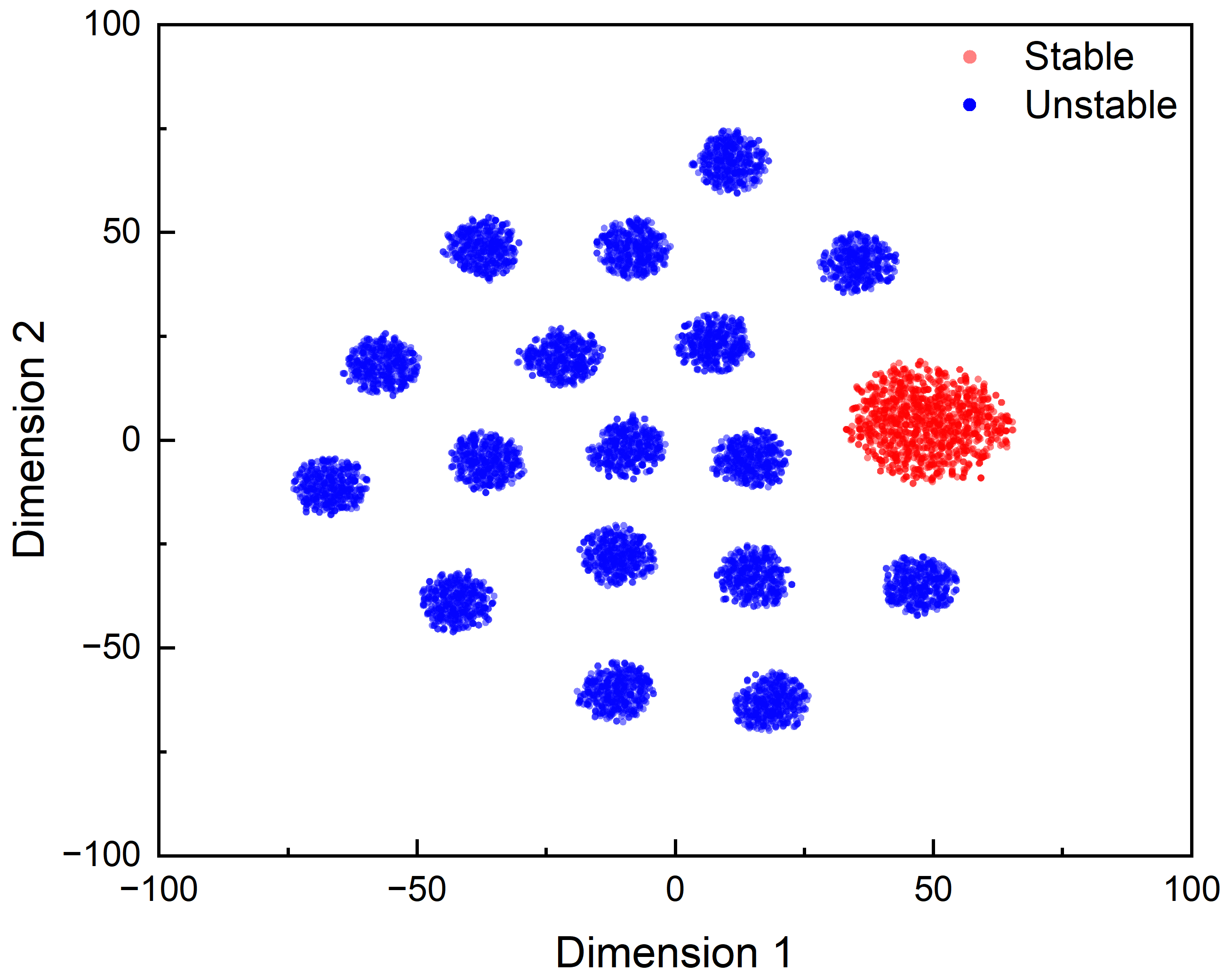}
         \caption{UC1 with only physical features.}
     \end{subfigure}
     \begin{subfigure}{0.45\textwidth}
         \centering
         \includegraphics[width=\textwidth]{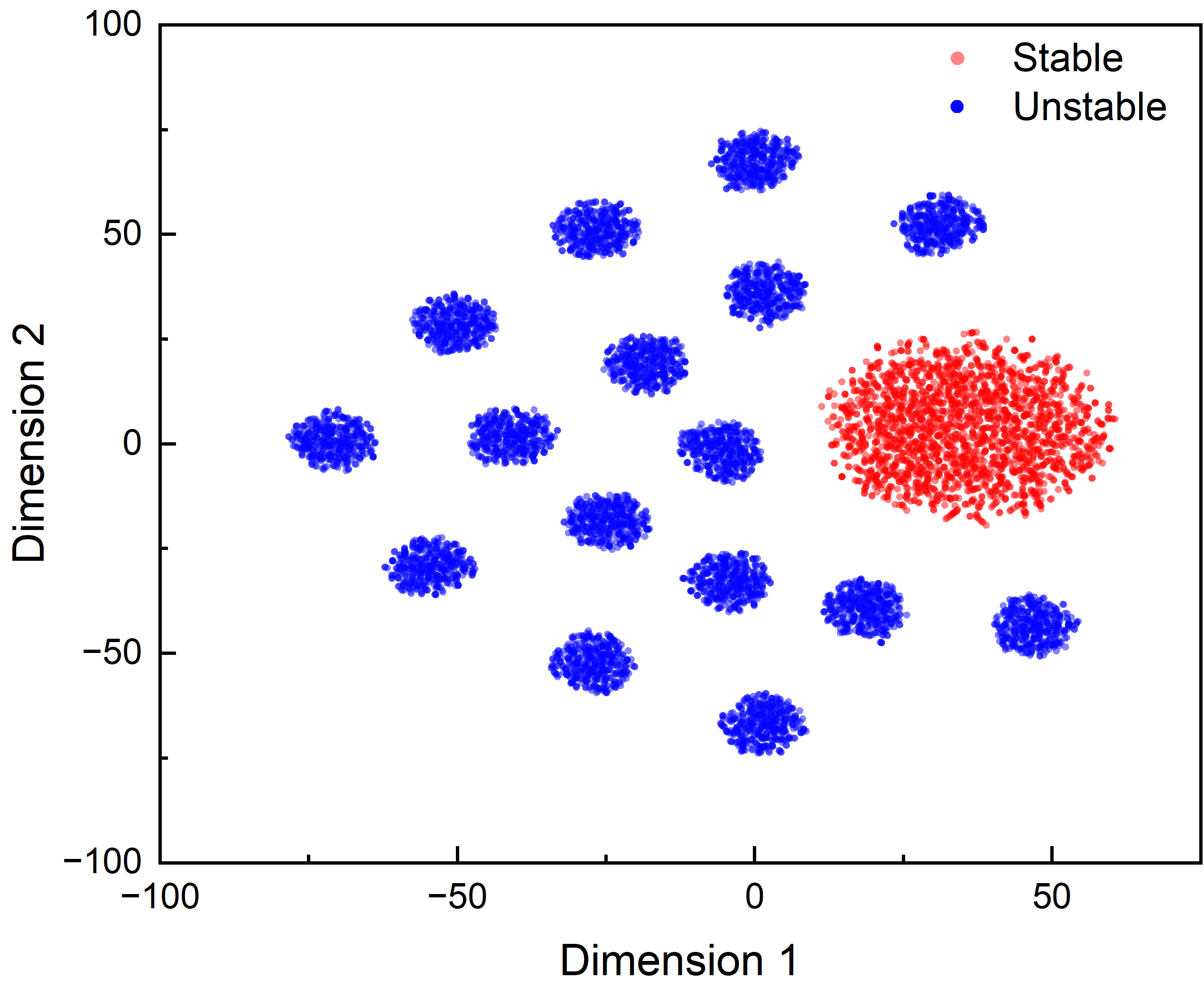}
         \caption{UC1 with cyber-physical features.}
     \end{subfigure} \hfill
     \begin{subfigure}{0.45\textwidth}
         \centering
    \includegraphics[width=\textwidth]{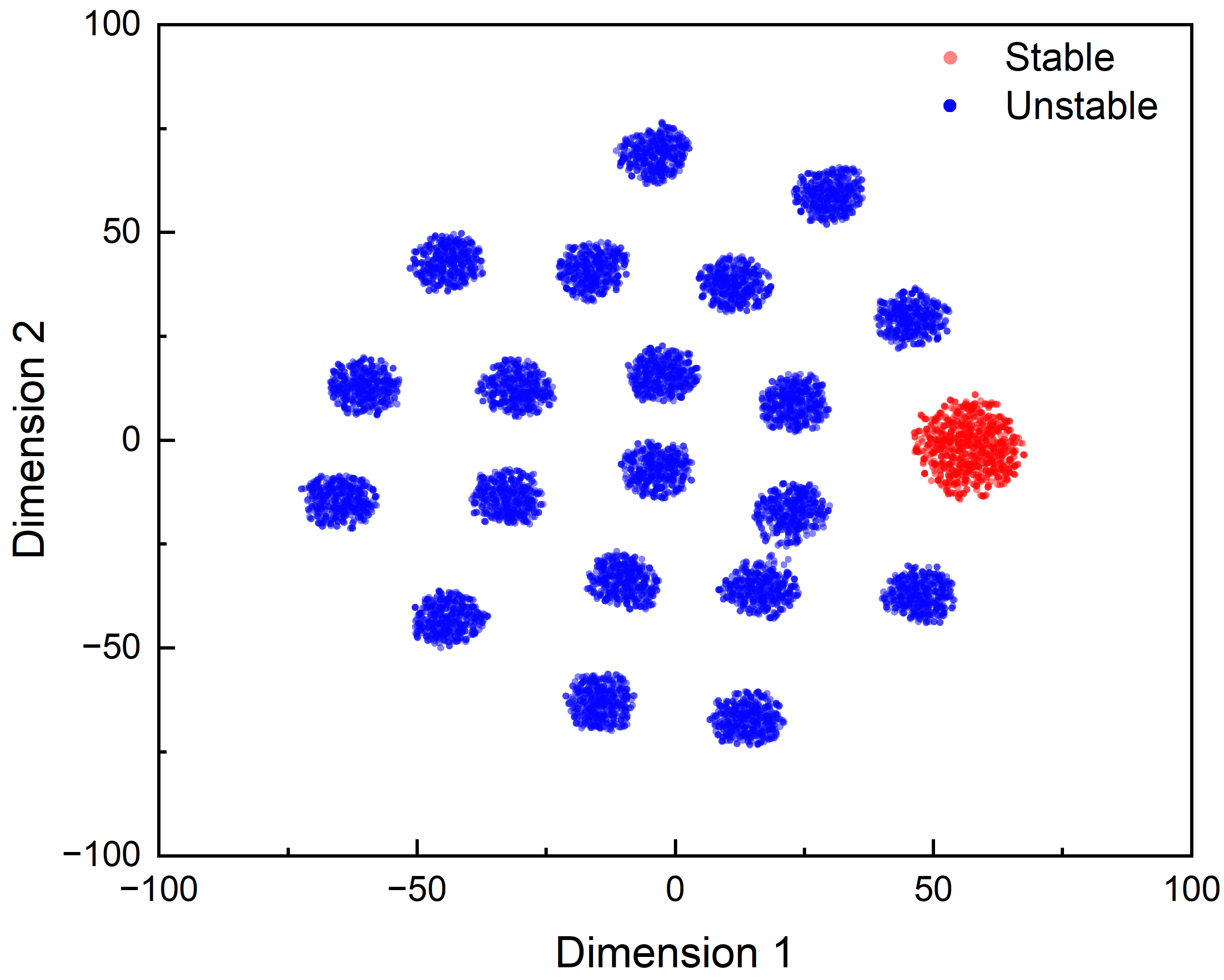}
    \caption{UC2 with only physical features.}
     \end{subfigure}
     \begin{subfigure}{0.45\textwidth}
         \centering
         \includegraphics[width=\textwidth]{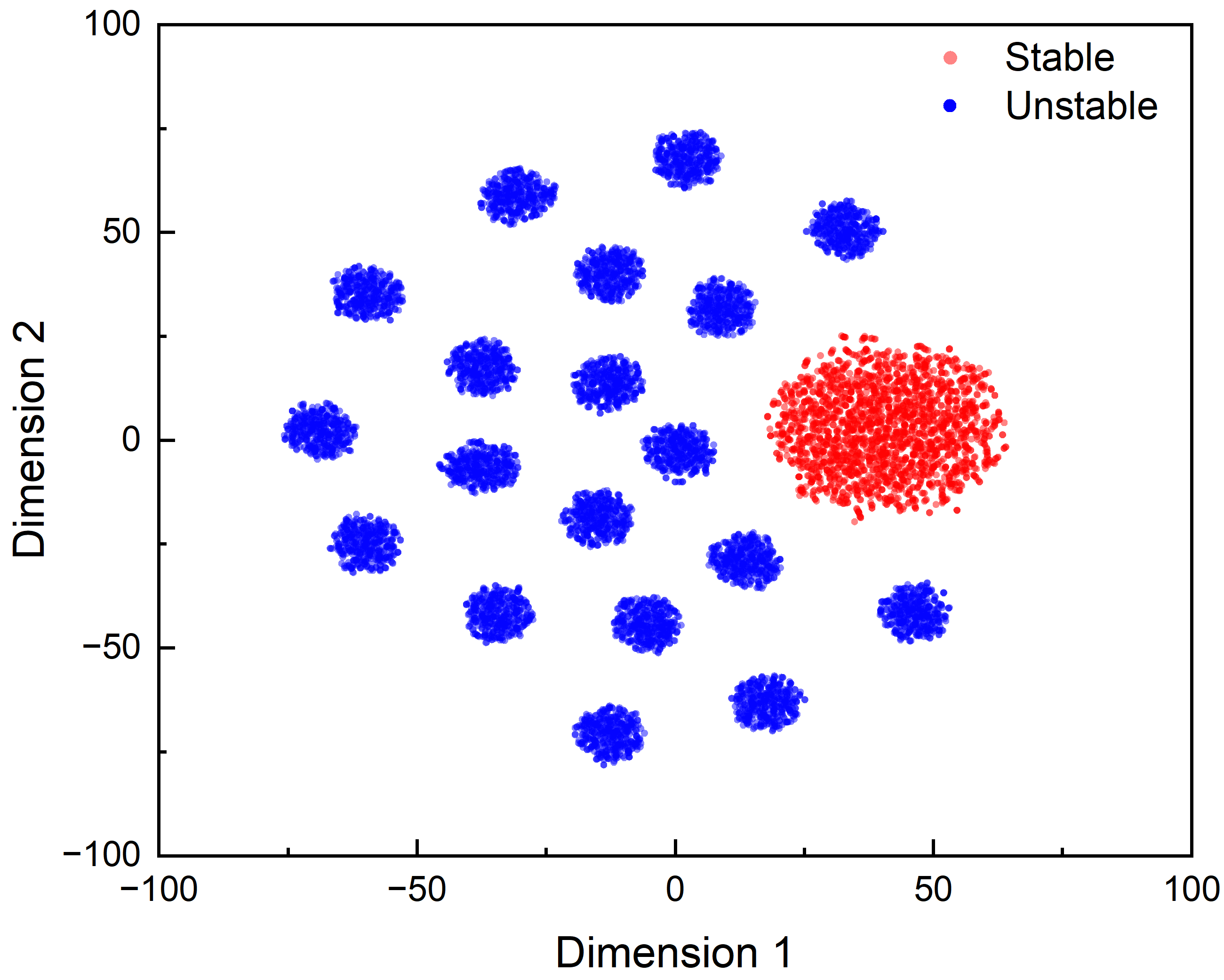}
         \caption{UC2 with cyber-physical features.}
     \end{subfigure}
    \caption{Clustering of IEEE 24-Bus System under DoS disturbance: (a) UC1 with only physical features, (b) UC1 with cyber-physical features, (c) UC2 with only physical features, and (d) UC2 with cyber-physical features, showing that the incorporation of cyber features contributes to the clarity with which one can identify a disturbance, as indicated by the separation between clusters. 
    }
    \label{fig:multimodal_figure}
\end{figure*}
\begin{figure}
  \includegraphics[scale=0.35]{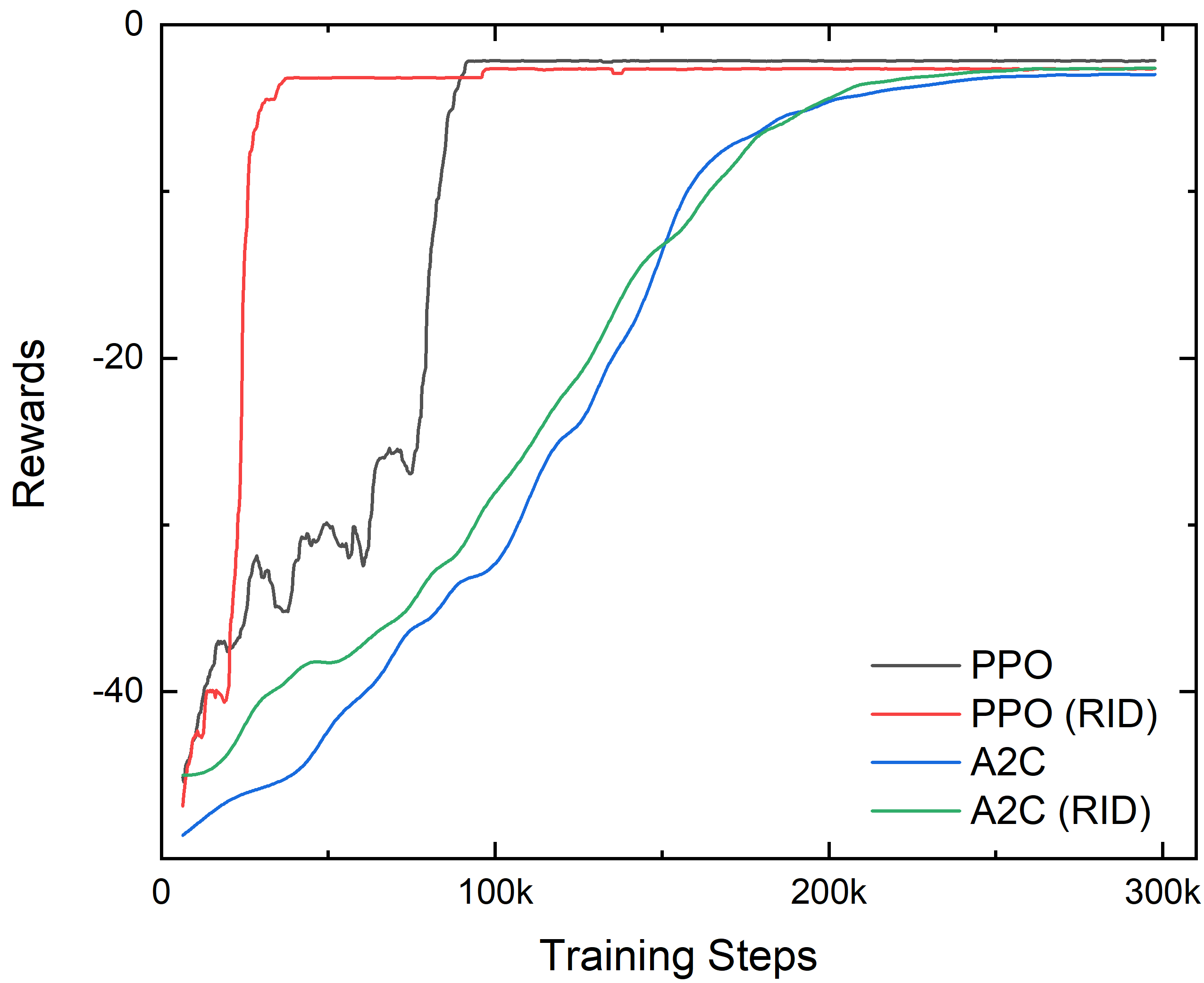}
\captionof{figure}{Training results for the IEEE 24 bus-system under UC2, with comparison of the PPO and the A2C with/without the RID module, showing that the integration of RID module significantly speeds up the training process of PPO, and slightly speeds up A2C.}
   \label{fig:24result1}
\end{figure}

In Fig. \ref{fig:multimodal_figure}, the stable and unstable data points are 
separated into different clusters. When cyber features are incorporated with physical features, the stable cluster increases in size and achieves better separation from the unstable cluster. Additionally, the inset of Fig. \ref{fig:multimodal_figure} displays the precision, recall, and $F_1$ score, respectively, providing the classification of disturbances. This performance is visually evident since the clusters are entirely separated with no overlap, meaning each data point is correctly classified and distinctly belongs to a particular cluster. As shown in Fig. \ref{fig:24result1}, the integration of the RID module significantly speeds up the training process of PPO by reducing the training steps of the RL agent while slightly speeding up the training process of A2C. The RID module reduces the action space of the RL agent by about 15-17\% for both cases. Besides, the rewards gained by PPO after applying RID are slightly higher than the one that applies PPO without RID, indicating a slight increase in loss. In contrast, the rewards for A2C after applying RID are slightly lower than without RID, resulting in reduced system loss.

\begin{table}[b]
\caption{RID Results for the augmented 24-bus system.}
\label{tab:RID_24bus}
\begin{tabular}{|l|l|l|}
\hline
\textbf{Scenarios} & \textbf{Essential Controllers} & \textbf{Critical Controllers} \\ \hline
UC1 & Batteries 2, 13 & Battery 2 \\ \hline
\begin{tabular}[c]{@{}l@{}} UC2 \end{tabular} & \begin{tabular}[c]{@{}l@{}}Capacitors 11, 6, 10, 15\\ Batteries 2, 13\end{tabular} & \begin{tabular}[c]{@{}l@{}}Capacitors 6, 10\\ No critical battery\end{tabular} \\ \hline
\end{tabular}%
\end{table}

\begin{figure}
\includegraphics[scale=0.70]{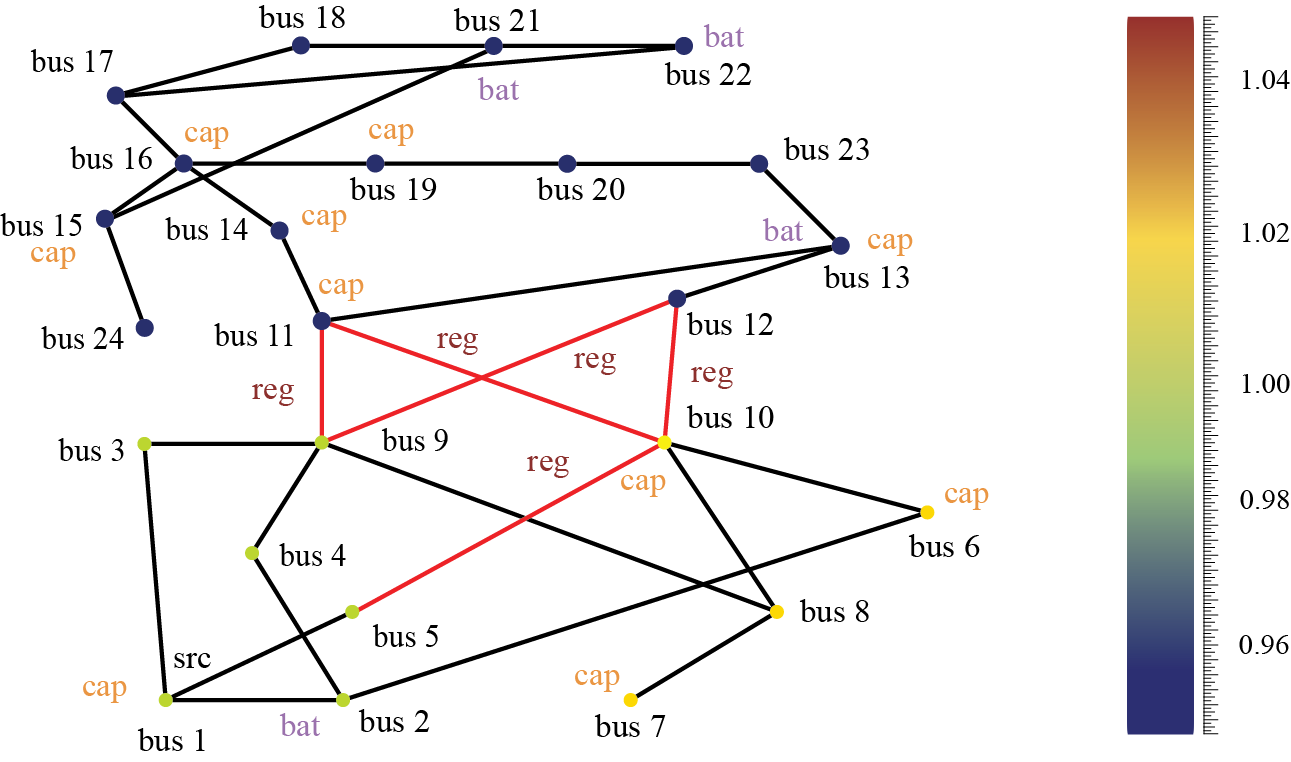}
\captionof{figure}{Bus voltages of the augmented IEEE 24-bus system presented via a heatmap (trained with PPO, where `cap', `bat', and `reg' denote capacitor, battery, and the tap of the transformer, respectively). All buses' voltages are kept within the ±5\% fluctuation.}
\label{fig:IEEE_24_heat_map}
\end{figure}
\begin{figure*}[!ht]
     \centering
     \begin{subfigure}{0.47\textwidth}
         \centering
         \includegraphics[width=\textwidth]{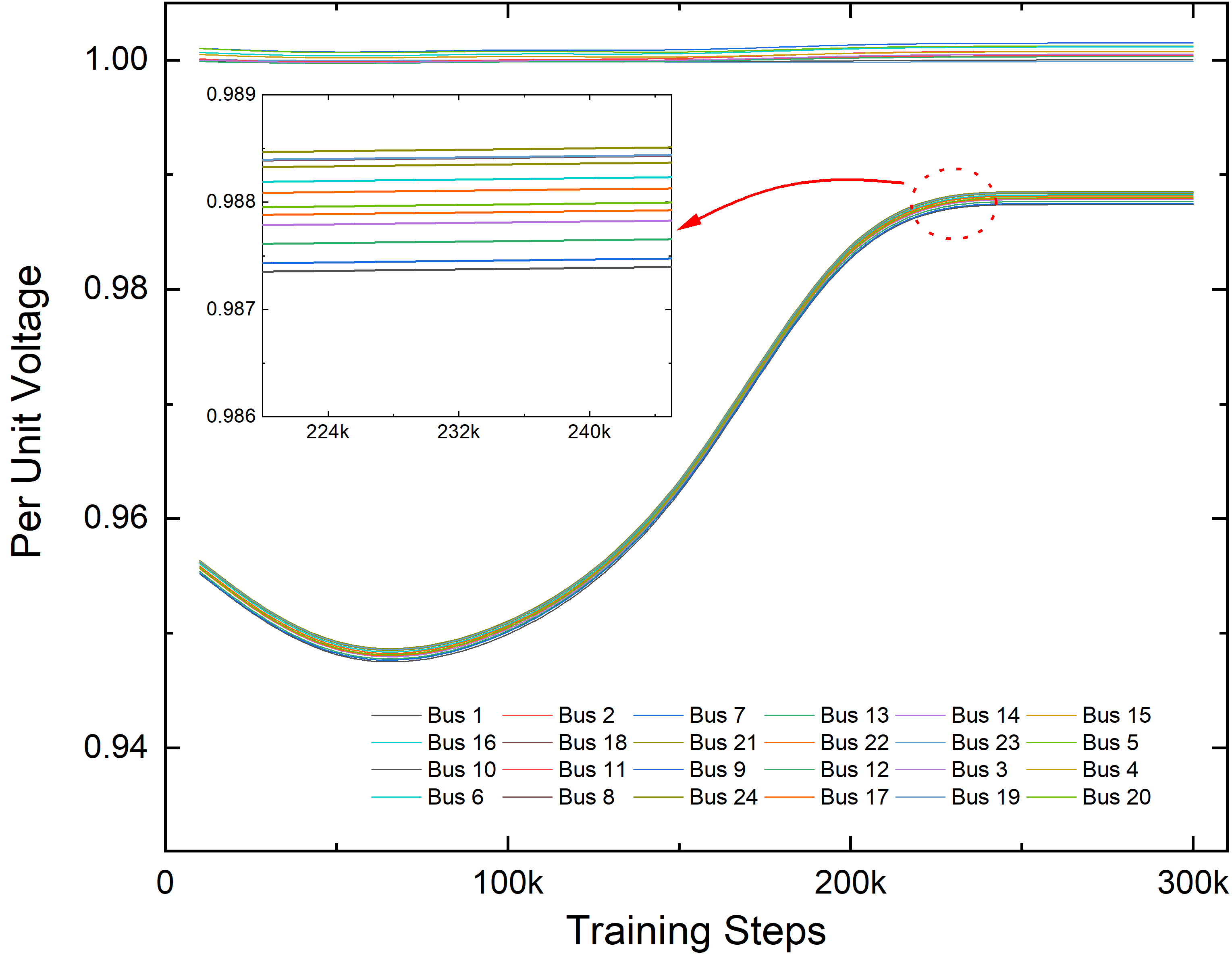}
         \caption{A2C without RID.}
     \end{subfigure}
     \begin{subfigure}{0.47\textwidth}
         \centering
         \includegraphics[width=\textwidth]{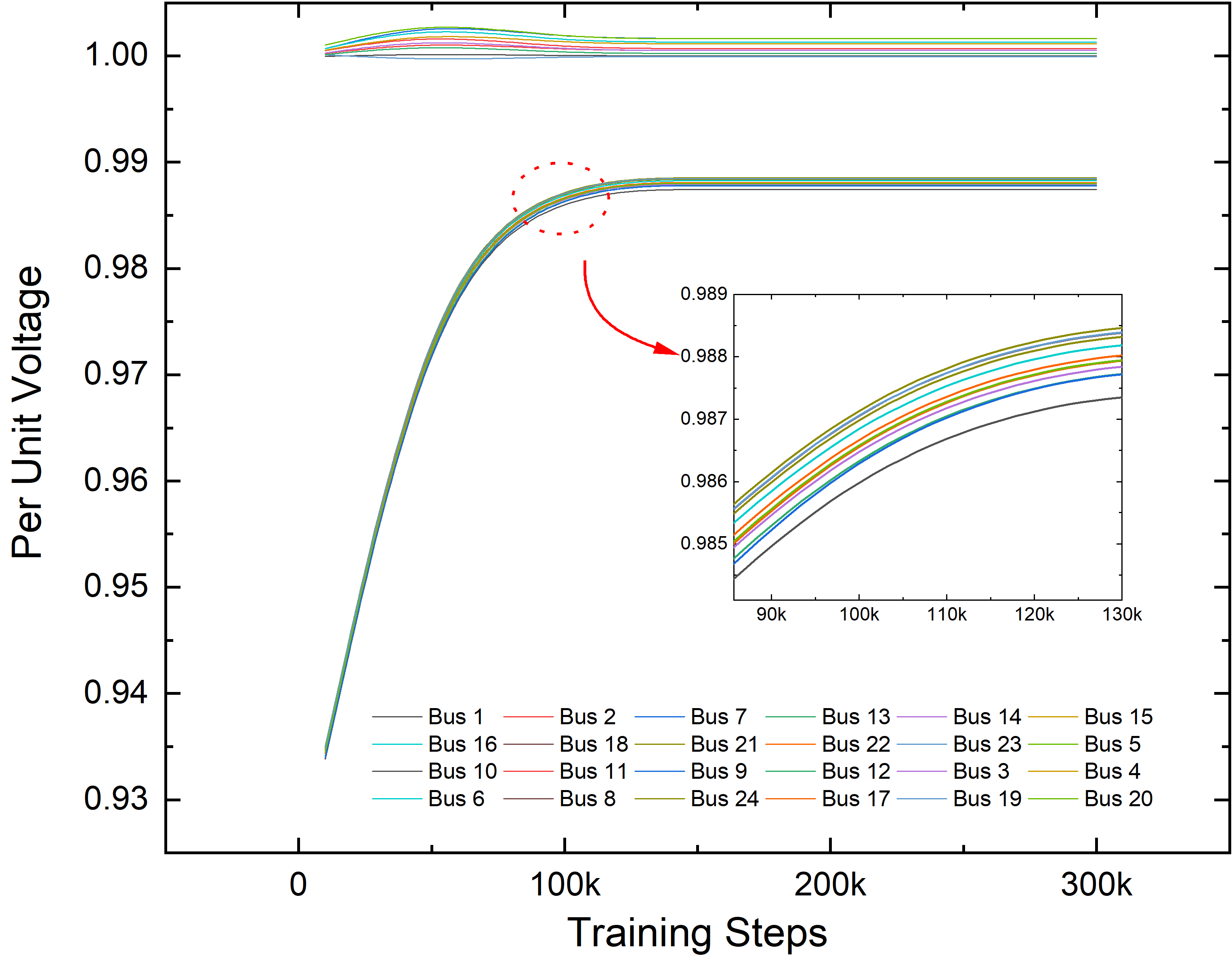}
         \caption{PPO without RID.}
     \end{subfigure} \hfill 
     \begin{subfigure}{0.47\textwidth}
     \vspace{4mm}
         \centering
    \includegraphics[width=0.99\textwidth]{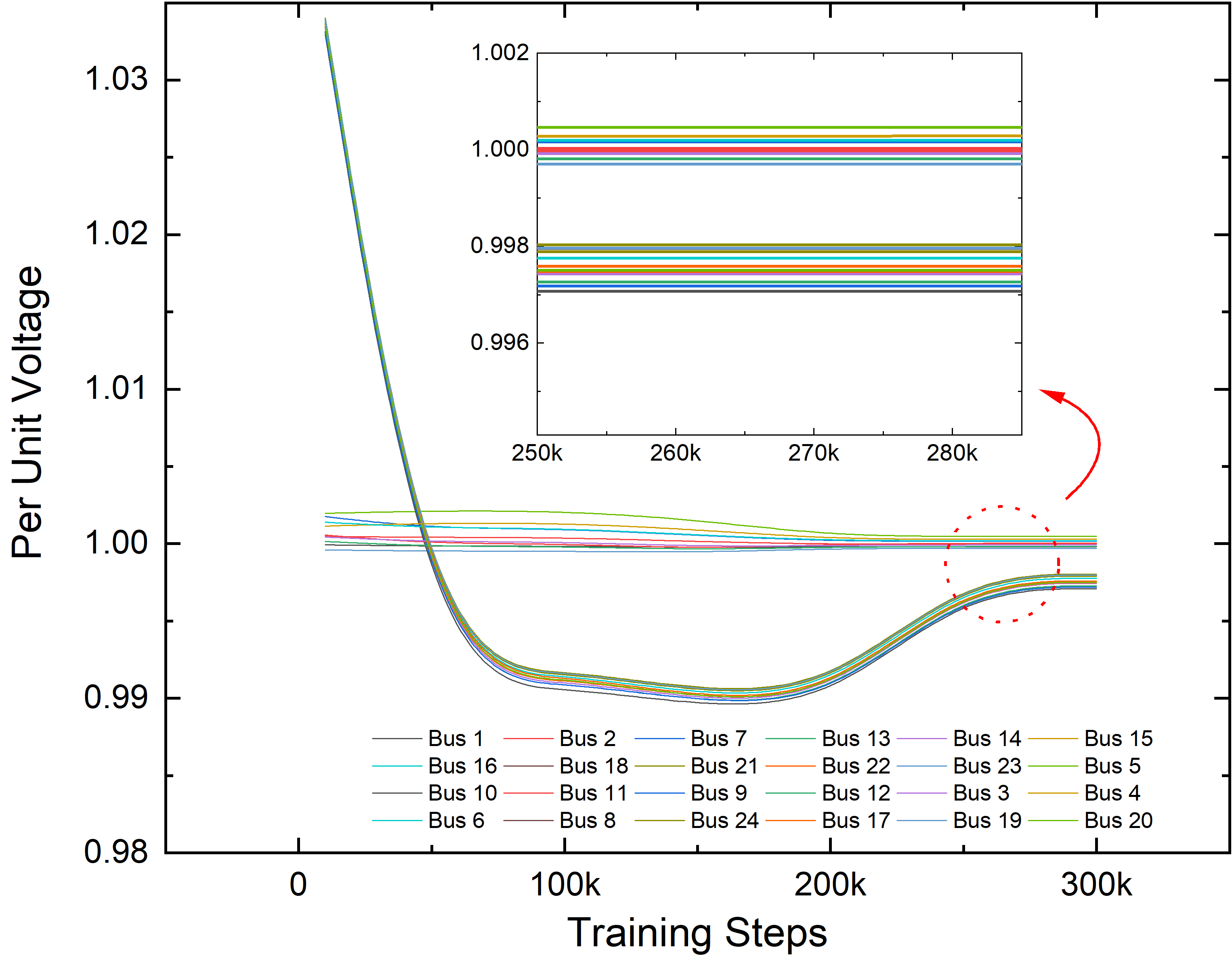}
    \caption{A2C with RID.}
     \end{subfigure}
     \begin{subfigure}{0.47\textwidth}
         \centering
         \includegraphics[width=\textwidth]{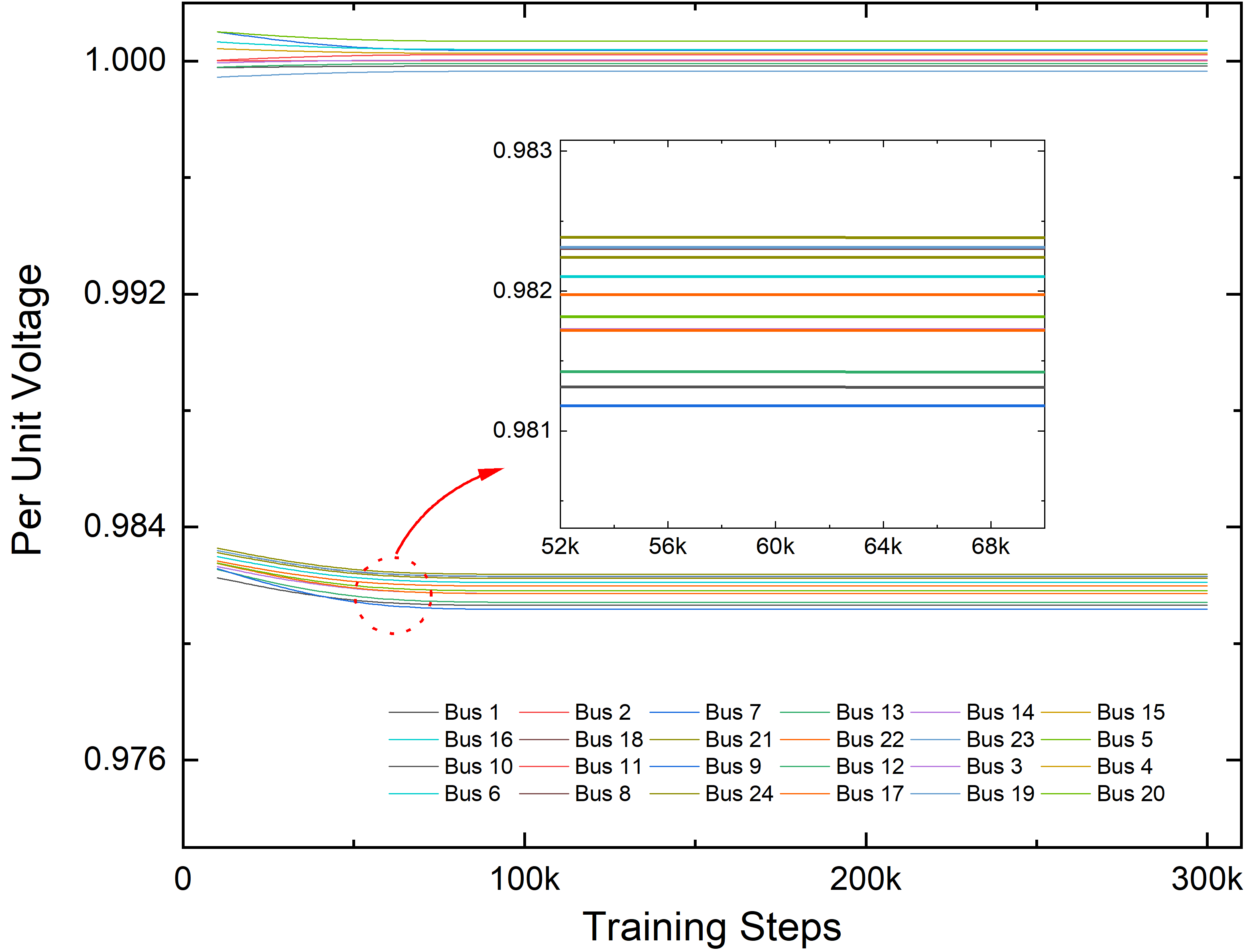}
         \caption{PPO with RID.}
     \end{subfigure}
        \caption{All bus voltages of IEEE 24-bus system under UC2 with/without the implementation of RID algorithm  (a) A2C without RID; (b) PPO without RID (c) A2C with RID; (d) PPO with RID. Close-up views reveal the stabilized voltages in steady states), showing that both PPO and A2C reached voltage steady states, but the PPO has  fewer fluctuations compared to the A2C. }
        \label{fig:24result2}
\end{figure*}

The bus voltages of the augmented IEEE 24-bus system are generated using PPO and represented using a heatmap shown in Fig.~\ref{fig:IEEE_24_heat_map}.  The heatmap illustrates the voltage levels across all buses within the system. By employing PPO in the response engine, all bus voltages are maintained within the voltage limits even during the DoS. In both UC1 and UC2, the PPO agents continue to outperform the A2C agents. Figs. \ref{fig:24result2}(a)-\ref{fig:24result2}(d) show the voltage variations at each bus during the training process. Detailed close-up views reveal the stabilized voltages in steady states for both PPO and A2C. Without using the RID algorithm, there are more voltage variations for the A2C as compared to PPO. When including the RID module, the voltage variations decrease once steady-state conditions are reached, particularly with the PPO agent. The experimental results prove that the developed GridResponder platform could effectively adjust the system's state back to normal states under a DoS attack. 


\section{Conclusion and Future Work}\label{sec:conclusion}

\subsection{Conclusion}

The RL-RID-GridResponder designed in this paper provides fast, accurate, and optimal responses under contingency, and it exhibits enhanced scalability by reducing the action and state spaces via RID and shows improved response by fusing cyber and physical data. It provides optimal management of grid-tied resources with the objectives of voltage regulation, control error reduction, and power loss minimization. By integrating RL techniques, GridResponder is capable of offering solutions with minimum loss to modern grid challenges and helping the grid adapt to real-time changes. In this paper, the engine is designed to regulate voltage levels across a large-scale grid. By optimizing control actions, the engine keeps the voltage profile stay within upper and lower bounds, 
and reduces energy losses, even under
DOS disturbances.
The RL-RID-GridResponder ensures the resilience of the grid, by 
helping the power system automatically learn to operate toward an optimal state. When under a DoS attack, the agents follow policy-based RL and regulate the bus voltages for the Volt-Var control problem using available grid resources. Simulation results corroborate the efficacy of the proposed optimal response engine on both the augmented WSCC 9-bus and the augmented IEEE 24-bus systems.

\subsection{Future Work}

Following the development of the RL-RID-GridResponder, several future research directions have been identified: 1) Lift the limitations on cyber-physical system environments. The RL bottleneck for verification and deployment is due in part to the lack of fused cyber-physical power system environments. Our next step includes the development of a more comprehensive cyber-physical environment for large-scale power system simulation; 2) RL models are often highly case-specific. State-of-the-art RL models tend to often only be trained and tested under a fairly limited set of scenarios. Therefore, future work could focus on training datasets that include a wide variety of cyber and physical disturbance scenarios. Additionally, methods to improve the model's adaptability, such as allowing the system to learn from previous experiences in different but related scenarios (\textit{transfer learning}), can enhance its generalization capabilities. These approaches could
further improve the resilience of power systems against unforeseen cyber-attack scenarios; 3) Another critical point in this area is the integration of various grid-tied resources within advanced CPS testbeds.  The interconnectivity and interoperability of different  grid-tied resources can potentially be advanced by designing such a scalable cyber-physical optimal response engine.



 
\section{Acknowledgement}\label{sec:ack}
The authors would like to acknowledge the  US Department of Energy 
under award DE-CR0000018, the National Science Foundation under Grant 2220347, and
the SCORE project team.

The authors would like to thank Sandia National Laboratories for supporting this work. 
Sandia National Laboratories is a multimission laboratory managed and operated by National Technology \& Engineering Solutions of Sandia, LLC, a wholly owned subsidiary of Honeywell International Inc., for the U.S. Department of Energy’s National Nuclear Security Administration under contract DE-NA0003525. This paper describes objective technical results and analysis. Any subjective views or opinions that might be expressed in the paper do not necessarily represent the views of the U.S. Department of Energy or the United States Government.

\bibliographystyle{IEEEtran}
\bibliography{ref.bib}
\end{document}